\documentclass[]{jfm}
\usepackage{amsmath}
\usepackage{graphicx}
\usepackage{booktabs}
\usepackage{makecell}
\usepackage{epstopdf,epsfig}
\usepackage{newtxtext}
\usepackage{newtxmath}
\usepackage{placeins}
\usepackage[normalem]{ulem}
\usepackage{natbib}
\usepackage{csquotes}
\usepackage{hyperref}
\usepackage{siunitx}
\usepackage{graphicx}
\usepackage{epstopdf, epsfig}
\usepackage{amsmath}
\usepackage{soul}
\usepackage{caption}
\usepackage{subcaption}
\usepackage{float}
\usepackage{color}
\usepackage{soul}
\usepackage{subcaption}
\usepackage{commath}
\usepackage{epstopdf, epsfig}
\usepackage{amsmath}
\usepackage{soul}
\usepackage{caption}
\usepackage{subcaption}
\usepackage{multirow}
\usepackage{xcolor}

\hypersetup{
    colorlinks = true,
    urlcolor   = blue,
    citecolor  = blue,
}
\usepackage{tikz}
\usepackage{pgfplots}
\pgfplotsset{compat=1.18}
\usetikzlibrary{external}
\newcommand{\akm}[1]{{\color{black}#1}}
\newcommand{\rkb}[1]{{\color{black}#1}}
\newcommand{\pfl}[1]{{\color{black}#1}}
\newcommand{\Frou}{\operatorname{\mbox{\textit{Fr}}}}

\definecolor{robin1}{rgb}{0.816,0.361,0.153}%

\shorttitle{An experimental and numerical study of the circular hydraulic jump}
\shortauthor{Mallik et al.}

\title{An experimental and numerical study of the circular hydraulic jump}

\author{Arnab Kumar Mallik$^1$, Hrvoje Jasak,$^2$, D. Ian Wilson$^3$, P. F. Linden$^1$, Rajesh K. Bhagat$^1$ \corresp{\email{rkb29@cam.ac.uk}},
 }

\affiliation{
$^1$Department of Applied Mathematics and Theoretical Physics, University of Cambridge, UK\\
$^2$Cavendish Laboratory, Department of Physics, University of Cambridge, UK\\
$^3$ Department of Chemical Engineering \& Biotechnology, University of Cambridge, UK}

\begin{document}
\maketitle

\begin{abstract}
This paper describes experiments and numerical simulations on the normal impact of a round liquid jet onto a horizontal surface, such as when water from a tap hits the bottom of a kitchen sink. In this case the liquid impacting on the surface spreads as a fast-flowing thin film and, at some distance from the point of impact, the thickness of the flow abruptly increases and the speed of the flow is reduced. In the experiments studied here the flow is axisymmetric about the axis of the jet, and the abrupt change in depth occurs at a given radius and is known as a circular hydraulic jump (CHJ). We present new experiments in which we measure the thickness of the liquid film inside and beyond the jump and use these measurements to estimate the governing parameters, the Weber, Froude and Reynolds numbers that determine the influence of surface tension, gravity and viscosity, respectively. We also carry out numerical simulations of the flow that show excellent agreement with the experiments and provide independent estimates of these dimensionless parameters. We find that, on the scale of a kitchen sink, and for water at high Reynolds number and low Bond number, at the jump the Weber number is of order unity while the Froude number is large, implying that the jump is controlled by surface tension. We also define a critical dimensionless jet flow rate at which this control no longer holds and gravity plays a significant role.
\end{abstract}

\section{Introduction}
\label{sec:intro}
The circular hydraulic jump, the abrupt thickening of a thin liquid film produced by a jet impacting  on a flat surface such as a kitchen sink, is among the most familiar of free-surface flows. These jumps have captivated many eminent thinkers, including Leonardo da Vinci \citep{marusic2021leonardo}. \pfl{In} the classical account \pfl{the jump  is controlled by gravity}, hence the use of the \pfl{term hydraulic} in the name of the jump. According to this account, close to the point of jet impact where the film is thin and fast the flow is supercritical, with speed faster than long gravity waves that would otherwise carry information upstream. Outside the jump the flow is subcritical with speed less than the long-wave speed. The jump is the transition between these two states, and  occurs where the flow speed, \(U\) equals the long-wave speed \(\sqrt{gh}\), where \(g\) is gravitational constant and \(h\) is the liquid layer thickness. Consequently, the location of the jump occurs  where the Froude number $\Frou = U/\sqrt{gh} = 1$. This description is well established for flows in which gravity provides the dominant restoring force \citep{rayleigh1914theory, craik1981circular, tani1949water}. \rkb{ By this account the jump is more than a local balance of forces: it is the point at which upstream wave propagation is arrested, so that disturbances generated in the subcritical film cannot penetrate the
supercritical interior \citep{jannes2011experimental,
rousseaux2020classical}. The gravitational theory of the circular jump was developed in a series of studies by Bohr and co-workers, combining shallow-water analysis with experiments over a range of flow rates and fluids \citep{bohr1993shallow, bohr1996hydraulic, bohr1997averaging}. }

However, later experiments by \cite{bhagat2018origin} on the normal impact of a jet onto plates at different orientations to the horizontal found that, for the same liquids and the same jet flow rates the radius of the jump was the same. Indeed, for the impact of a horizontal jet onto a vertical plate the resulting jump was almost exactly circular despite the direction of gravity opposing the flow at the top and assisting the flow at the bottom of the jump region. These observations naturally raise the question of the role of gravity in these jumps. This is the question we will address in this paper.
 
The paper is organised as follows. We begin in \S\ref{sec:scaling} with a scaling analysis that sets out the relevant dimensionless parameters describing the flow. The experiments are described in \S\ref{sec:experiment}, and the measurements are presented in \S\ref{Sec: experimental observation}: the evolution of the film and the jump in \S\ref{ssec:qual}, the measured film profiles in \S\ref{ssec:quant}, and the local Bond, Froude and Weber numbers obtained from them in \S\ref{sec:Bo_Fr_We}, which together establish the regime in which the jump forms. 
The flow is examined numerically in \S\ref{sec:numerics}. Simulated and measured profiles are compared through the transient \pfl{development of the flow in } \S\ref{sec:Transient state} and at steady state in \S\ref{sec: steady state}. Two \pfl{further} computations \pfl{examine the jump dynamics} directly: the surface tension is varied alone, at fixed density, viscosity and flow rate in \S\ref{sec:surface_tension}, and the wetting condition is varied over the full range of contact angles in \S\ref{sec:contact_angle}.  The conclusions follow in \S\ref{sec:conclusion}.  
\begin{figure}
\centering
\includegraphics [width=1\linewidth]{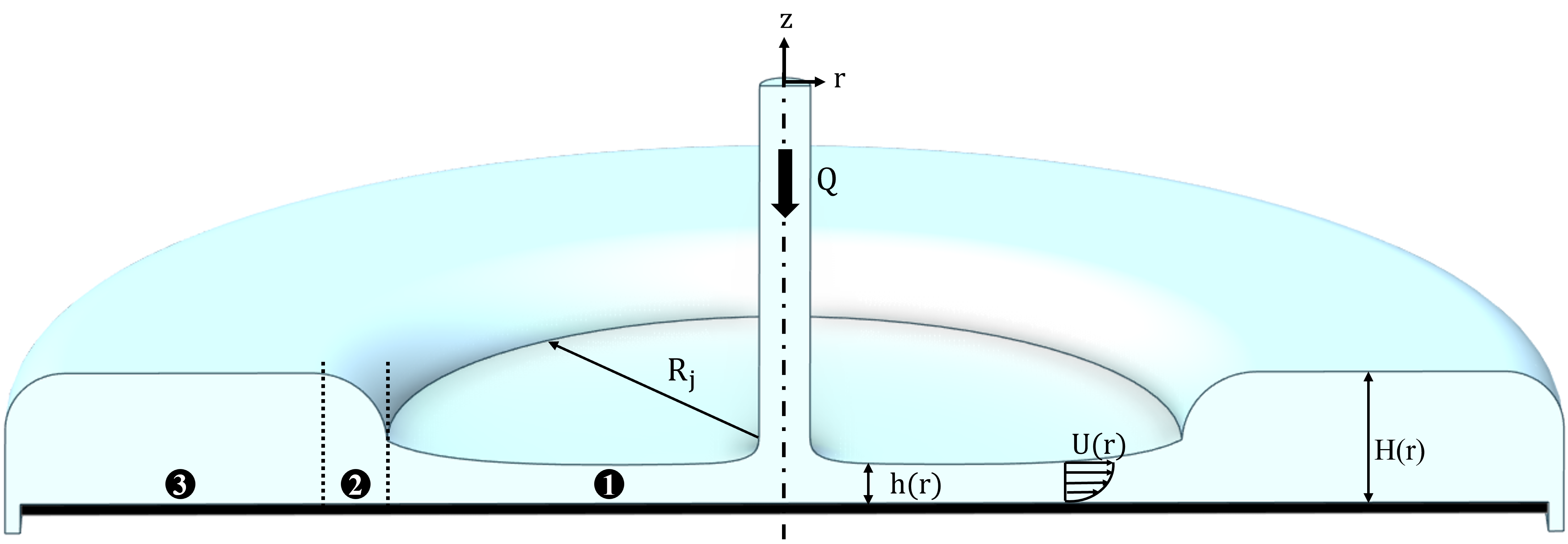}
\caption{ A schematic of the \pfl{flow and the} circular hydraulic jump \pfl{produced by the normal impact of a jet onto a planar surface} . Label 1 denotes the supercritical thin-film region \pfl{of depth $h(r)$ and depth-average radial velocity $U(r)$} , label 3 denotes the subcritical thick-film region \pfl{of depth H(r)}, and label 2 marks the \pfl{jump between these regions, of radius} $R_j$.   }
\label{fig: CHJ schematic}
\end{figure}

\section{Scaling analysis}
\label{sec:scaling}
In this section, dimensional analysis is used to identify a minimal set of dimensionless parameters governing the circular hydraulic jump. For a steady axisymmetric flow, the radius $R$ of the jump is determined  as a function of eight parameters
\begin{equation}
    R=\mathcal{F}(h, U, d , Q, \rho,\nu,\gamma,g),
\end{equation}
where $h$ is the film thickness, $U$ is the (radial) flow speed, $d$ is the jet diameter, 
$Q$ is the jet flow rate, $g$ is the acceleration due to gravity, and $\rho$, $\nu$ and $\gamma$ are the density, kinematic viscosity and surface tension of the liquid, respectively. 
The problem contains nine dimensional quantities involving the three fundamental dimensions mass, length and time. Buckingham's $\Pi$-theorem therefore gives six \pfl{independent} dimensionless groups given by 
\begin{equation*}
\Pi_h=\frac{h}{R},
    \qquad
\Pi_d=\frac{d}{R},
    \qquad
\Pi_\nu = \frac{Uh}{\nu},
\qquad
\Pi_\gamma = \frac{\rho U^2 h}{\gamma},
\qquad
\Pi_g = \frac{U^2}{gh},
\qquad
\Pi_Q = \frac{2 \pi\gamma^2}{Qg\rho^2\nu},
\end{equation*}
\pfl{where the factor $2\pi$ in the last group is added for convenience.}

The groups $\Pi_\nu$, $\Pi_\gamma$ and $\Pi_g$ represent local measures of the relative importance of viscosity, surface tension and gravity, and can be identified as Reynolds ($Re$), Weber ($We$) and \pfl{squared} Froude ($Fr^2$) numbers, respectively.  
Previous experiments indicate that,  provided $d \ll R$ which is the case here, the jump radius is only weakly affected by the jet diameter $d$ \citep{stevens1993measurements, choo2016influence, bhagat2018origin}. Consequently, we ignore $\Pi_d$ and the dimensionless film thickness may be expressed as
\begin{equation}
    \frac{h}{R}
    = \mathcal{F}\left(
        Re, We, Fr^2, \Pi_Q
    \right).
\end{equation}

\rkb{We first resolve the status of $\Pi_Q$. \pfl{Taking $U$ as the depth-averaged radial veloocity} the axisymmetric volume flux is
$Q = 2\pi r h U$. It is convenient to introduce the volume flux per radian
$q\equiv Q/(2\pi)$, so that mass conservation reads $q = rhU$. Applying this to the
group definitions gives
\begin{equation}\label{eq:Qstar_groups}
\Pi_Q = \frac{2\pi\gamma^{2}}{Q\rho^{2}\nu g} = \frac{\gamma^{2}}{q\rho^{2}\nu g}\equiv \frac{h}{R}\,\frac{Re\,Fr^{2}}{We^{2}}\equiv Q^{*},
\end{equation}
so that $\Pi_Q$ is fixed by the remaining groups through continuity. 
\pfl{We discuss the role of $Q^*$ below but for the moment} eliminating $Q^{*}$, the dimensionless film thickness reduces
to
\begin{equation}\label{eq:response_reduced}
    \frac{h}{R} = \mathcal{F}\!\left(Re,\,We,\,Fr^{2}\right).
\end{equation}
Combining $h/R$ and $Re$ with \pfl{limits set by values of either} $We$ or $Fr^{2}$ recovers the two extreme
scalings of the jump radius, as we now show.
 
For a thin film, conservation of mass together with this inertia--viscous
balance imply
\begin{equation} \label{eq:mm}
    U R h \sim q, \qquad
    \frac{U^{2}}{R} \sim \nu\frac{U}{h^2}.
\end{equation}

Rearranging
\eqref{eq:mm} isolates the Reynolds number,
\begin{equation}\label{eq:hR_Re}
    \frac{h}{R} \sim \frac{\nu}{U\,h} = \frac{1}{Re},
\end{equation}
\pfl{which for a thin film $h \ll R$ implies $Re \gg 1$}.  Using continuity, we get the film relations
\begin{equation}\label{eq:film}
    h \sim \frac{\nu R^{2}}{q}, \qquad U \sim \frac{q^{2}}{\nu R^{3}}
    \qquad\bigl(\text{equivalently } U\,h\sim q/R\bigr),
\end{equation}
which we use to close each of the two jump scalings with a single substitution.
 
Consider first a jump controlled by gravity, so that $Fr^2\sim 1$ at the jump,
\begin{equation}
    U^{2} \sim gh .
\end{equation}
Substituting the film relations \eqref{eq:film} gives
$q^4/(\nu^2R^6)\sim g\,\nu R^2/q$, hence $R^{8}\sim q^{5}/(g\nu^{3})$ and we get the gravity radial scale
\begin{equation} \label{eq:Rg}
    R_g = \frac{q^{5/8}}{(\nu^{3} g)^{1/8}},
\end{equation}
where the subscript $g$ denotes \pfl{that the Froude number condition invokes} the role of gravity. Alternatively, for a jump controlled by surface tension, $We\sim 1$ at the jump,
\begin{equation}
    \frac{\rho U^{2} h}{\gamma} \sim 1 .
\end{equation}
Substituting \eqref{eq:film} gives $\rho q^{3}/(\nu R^{4})\sim\gamma$, hence the capillary radial scale is
\begin{equation} \label{eq:Rst}
    R_{\gamma} = \left(\frac{\rho q^3}{\nu \gamma}\right)^{1/4},
\end{equation}
where here the subscript $\gamma$ denotes the role of surface tension.
 
Note the similar power-law dependence on the jet flow rate for the two scalings,
$R_g\propto Q^{5/8}$ and $R_\gamma\propto Q^{3/4}$. Flow rate is the parameter
most readily varied in experiments, yet the closeness of these exponents makes
it hard to discriminate between the two mechanisms from the jump radius alone. It
is therefore important to have additional, local measures of the relative
importance of gravity and surface tension, as developed below.
 
The two scalings coincide, $R_g=R_\gamma$, at the critical flux
$q_c=\gamma^{2}/\rho^{2}\nu g$, i.e. at the critical total flow rate
\begin{equation}\label{eq:Qc}
    Q_c = 2\pi q_c = \frac{2\pi\gamma^2}{\rho^2\nu g},
\end{equation}
which depends only on the liquid properties and delineates the relative
importance of surface tension and gravity. The corresponding dimensionless
flow-rate parameter
\begin{equation}\label{eq:Qstar}
    Q^{*} = \frac{Q_c}{Q} = \frac{q_c}{q} = \frac{2\pi\gamma^2}{Q\rho^2\nu g},
\end{equation}
is precisely $\Pi_Q$; here it acquires its physical
meaning, separating the regimes
\begin{equation}\label{eq:regimes}
   Q^{*} \begin{cases}
      \gg 1, & \text{capillary-dominated regime},\\[2pt]
      \approx 1,  & \text{mixed regime},\\[2pt]
      \ll 1, & \text{gravity-influenced regime}.
      \end{cases}
\end{equation}
More generally, the two candidate radii stand in the ratio
\begin{equation}\label{eq:Rg_Rgamma}
    \frac{R_g}{R_\gamma} = {Q^{*}}^{1/8},
\end{equation}
which reduces to $R_g=R_\gamma$ at $Q^{*}=1$ and orders the regimes of
\eqref{eq:regimes}. Since both $We$ and $Fr^{2}$ \rkb{decrease monotonically
with radius through the viscous film, the transition is triggered by
whichever criticality condition is met first,} i.e. at the smaller of the
two radii: $Q^{*}>1$ gives $R_\gamma<R_g$ and a capillary-controlled jump,
whereas $Q^{*}<1$ gives $R_g<R_\gamma$ and a gravity-controlled jump. The weak,
one-eighth-power dependence in \eqref{eq:Rg_Rgamma} keeps the two candidate radii
within an $O(1)$ factor of one another under any variation of flow rate, so
neither the flow-rate scaling nor a direct comparison of the measured radius
with $R_g$ or $R_\gamma$ cleanly identifies the mechanism. The dependence on the
fluid properties does discriminate. In particular, $R_\gamma\propto\gamma^{-1/4}$
whereas $R_g$ is independent of surface tension: a capillary-controlled jump
expands as $\gamma$ is lowered, while a gravity-controlled jump does not respond
to changes in $\gamma$. \pfl{We will return to this point when discussing the experimental and numerical results below.} 
 
Finally, the relative importance of gravity and surface tension follows directly
from the $\Pi$ groups as the Bond number,
\begin{equation} \label{eq:Bo}
    Bo = \frac{\Pi_\gamma}{\Pi_g} = \frac{We}{Fr^{2}}
       = \left(\frac{h}{l_c}\right)^{2},
    \qquad l_c \equiv\sqrt{\frac{\gamma}{\rho g}},
\end{equation}
where $l_c$ is the capillary length. The relevant comparison is thus between
$We$ and $Fr^{2}$---not $Fr$---since each is a ratio of inertia to a restoring
pressure (capillary for $We$, hydrostatic for $Fr^{2}$), so their quotient is
exactly $Bo$ and measures the relative importance of the two forces directly.
Measurements of the local film thickness and the application of mass continuity yield
the local \pfl{depth-averaged} velocity $U$, and hence the local $Bo$, $We$ and $Fr^{2}$, and shows how
they vary throughout the flow and, in particular, across the jump. The
experiments described in the next section are consequently designed to make
accurate measurements of the flow depth. This paper \pfl{concentrates on} the
regime $Q^{*}>1$, the capillary-controlled regime.}

\section{Experimental methods}
\label{sec:experiment}
The experiments were designed to quantify the liquid-film thickness in the supercritical and subcritical regions of the circular hydraulic jump and to investigate the influence of the contact line on the jump radius. Three working liquids were used: \pfl{pure} water, a water–glycerine solution (WG 8/7) mixed in an 8:7 volume ratio, and a water–2-propanol solution (W2P 95/5) mixed in a 95:5 mass ratio. The physical properties of these liquids are listed in Table~\ref{tab:fluid properties}. 
\par

\begin{figure}
\centering
\includegraphics[width=0.8\linewidth]{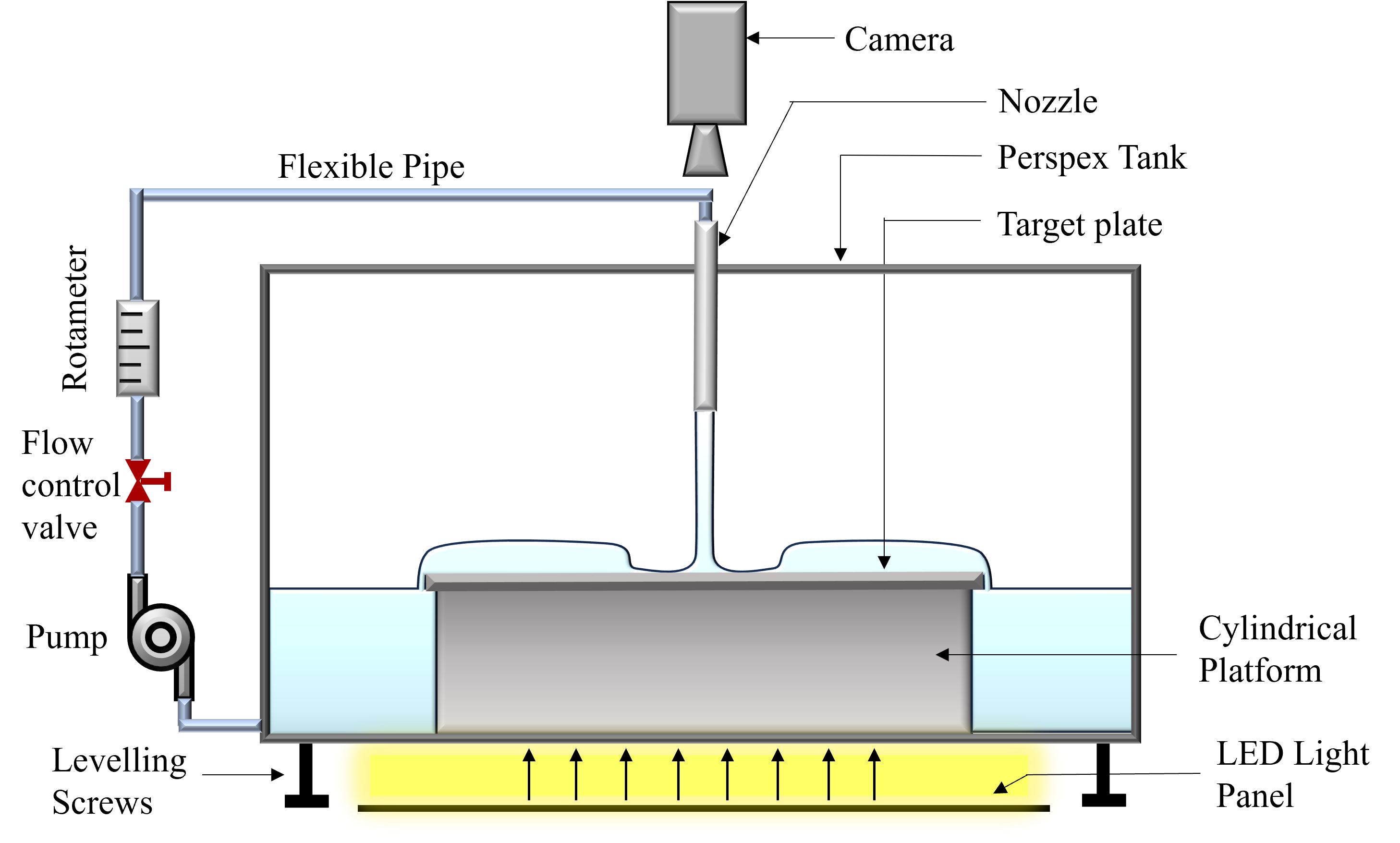}

  \caption{(a) Schematic of the hydraulic jump experiment. The \pfl{transparent} target plate \pfl{is mounted on a cylindrical platform and} is illuminated by a LED panel installed under the tank. A camera is positioned above to capture the top view of the experiment. Liquid - nigrosin solution is used as a working fluid \pfl{and is recirculated by the pump, keeping a constant liquid level in the tank}. } 
  \label{Fig: experimental setup}
\end{figure}

The liquid jets were directed vertically onto a horizontal plate mounted on a cylindrical platform \pfl{placed} within an acrylic tank of dimensions ($\SI{0.65}{} \times \SI{0.65}{} \times \SI{0.45}{\metre}$), which also acted as the liquid reservoir. The experimental setup is shown in figure \ref{Fig: experimental setup}. The tank was fitted with adjustable levelling supports, allowing the target plate to be levelled precisely so that the liquid film spread uniformly in all radial directions before reaching the edge of the plate. The liquid level in the reservoir was maintained approximately $\SI{2}{\milli\metre}$ below the upper surface of the target plate. This ensured that the \pfl{liquid falling from the edge of the plate} merged smoothly with the surrounding liquid pool, thereby avoiding the formation of a contact line \pfl{or any discontinuity of the free surface slope} at the plate edge. The liquid jet was generated using a centrifugal pump, which drew liquid from the reservoir, thereby maintaining a constant fluid depth in the reservoir, and delivered it through a nozzle of diameter $\SI{3}{\milli\metre}$ onto the target plate \pfl{(figure~\ref{Fig: experimental setup})} .   The volume flow rate $Q$ was controlled using a flow-control valve and measured with a calibrated rotameter. \pfl{The flow rate was constant during the experiment and known with a precision of \SI{0.01}{\liter \per \minute}.} \akm{Small fluctuations in the pump speed introduced slight oscillations in the flow rate. These fluctuations are accounted for later in this section as part of the uncertainty in the measured jump radius.}
\par
A circular transparent Perspex plate of diameter $\SI{240}{\milli\metre}$ was used as the target surface and was illuminated uniformly from below \pfl{by a LED light panel}. The working fluid was a nigrosin--\pfl{liquid} solution with a dye concentration of $\SI{0.1}{\gram\per\litre}$; the nigrosin dye rendered the \pfl{liquid} dark blue and enabled film-thickness measurements using light absorption, described in detail in Appendix \ref{append: dye calibration}. The method provided a continuous calibration curve and enabled rapid calibration at the start of each experiment. A similar approach was used by \citet{craik1981circular} to measure the film thickness of water in the supercritical region. Images \pfl{of the flow} were recorded from above using a monochrome Teledyne DALSA camera with a 4-megapixel resolution at a frame rate of 100 frames per second. MATLAB was used to compute the intensity ratio $I(t)/I_0$, where $I_0$ is the initial intensity recorded before the experiment started. These images were processed using ImageJ to determine the film thickness, and  the depth of the liquid layer was determined with a precision of \SI{0.016}{\milli \meter}. \akm{The jump location was determined directly from the measured film-thickness profiles. The transition is characterised by a rapid increase in $h$, and the jump location was therefore identified from the onset of a large positive gradient in the smoothed thickness profile. A threshold value of $dh/dr > 0.02$ was used for this purpose. The noise in the measured data from the camera sensor and illumination was smoothed using a Savitzky–Golay filter \citep{savitzky1964smoothing,schafer2011savitzky}. The filter parameters were selected based on the characteristics of the background noise and are discussed in detail in Appendix \ref{append: data smoothing}.  }

 \begin{table}
\caption{Fluid properties and flow-rate ranges used in the experiments.}
\label{tab:fluid properties}
\centering
\begin{tabular}{ccccccc}
\hline
\textbf{Fluid} & \textbf{$\rho$ ($\SI{}{\kilogram \meter^{-3}}$)} & \textbf{$\nu$ ($\SI{}{\meter^2 \second^{-1}}$)} & \textbf{$\gamma$ ($\SI{}{\newton \meter^{-1}}$)}& Q $(\SI{}{\liter \per \minute})$ & \textbf{$Q^*$} & \makecell{Reference for \\ fluid properties} \\
\hline
Water     & $1000$ & $1.02 \times  10^{-6}$ & 0.072 & 1.1 - 2.6 & 75.1--177.6 & - - \\

\makecell{Water Glycerine \\ (WG 8/7)}      & $1132$ & $6 \times  10^{-6}$ & 0.067 & 0.6 - 1.9 & 11.8--37.4 & \makecell{\cite{cheng2008formula};\\ \cite{volk2018density}; \\ \cite{takamura2012physical}}\\ 

\makecell{Water 2-Propanol\\ (W2P 95/5)} & $989$ & $1.13 \times 10^{-6}$ & 0.0496 & 0.86 - 2 & 42.8 -- 99.5 & \makecell{\cite{pang2007densities};\\ \cite{vazquez1995surface}} \\ 
\end{tabular}
\end{table}
The circular hydraulic jump experiments are relatively straightforward, and careful operation of the experimental setup does not introduce any appreciable error. Nevertheless, some unavoidable uncertainties remain in the measurements of the jump radius and film height, which are estimated here. To assess the repeatability of the jump-radius measurements, selected flow rates were repeated, namely \SI{2}{\liter\per\minute} for water and \SI{1.5}{\liter\per\minute} for the WG 8/7. In both cases, the variation in the measured jump radius was $< 1\%$. The fluctuations in the flow rate generated by the pump cause oscillations in the jump radius. To quantify this error, ten images were analysed at different times after the jump had reached a steady state. We evaluated the film profiles at ten different azimuthal angles over $360^{\circ}$ and found the \pfl{film thickness and the location of the} hydraulic jump to remain axisymmetric. The maximum relative uncertainty in the measured jump radius was found to be $\pm 6\%$, corresponding to an absolute variation of $\pm\SI{3.06}{\milli\metre}$ at the maximum flow rate of $\SI{2.6}{\litre\per\minute}$. The film height measurement may be affected by several sources of uncertainty, whose combined contribution is estimated using the uncertainty-propagation method described by \citet{moffat1988describing} and discussed further in appendix \ref{append: data smoothing}. The uncertainty in the film-height measurement is estimated to be $\pm 3.37\%$. The corresponding maximum and minimum absolute uncertainties are presented in table~\ref{tab:experimental_uncertainty}.

\begin{table}
\centering
\caption{ Experimental uncertainty in the measurements of jump radius and film height.}
\label{tab:experimental_uncertainty}

\small
\setlength{\tabcolsep}{3pt}

{
\begin{tabular}{ccccccccc}
\toprule
Quantity &
\makecell{Source of\\uncertainty} &
\makecell{Error /\\uncertainty } &
\makecell{Max. absolute \\uncertainty \\ (mm)} &
\makecell{Min. absolute\\uncertainty \\(mm)} \\
\midrule

$R_j$
&
\makecell{Pump \\fluctuation}
&
$\pm 6\%$
&
$\pm 3.06$
&
$\pm 0.42$
\\
\\
$h$
&
\makecell{Dye mass,\\ $l_w$,\\$h_w$}
&
$\pm 3.37 \%$
&
\makecell{$ \pm 0.128$ \\ (subcritical region)}
&
\makecell{$ \pm 0.0084$ \\ (supercritical region)}
\\
\end{tabular}
}
\end{table}

\section{Experimental Observations}
\label{Sec: experimental observation}

\subsection{Qualitative observations} \label{ssec:qual}

\begin{figure}
    \centering
    \includegraphics[width=0.8\linewidth]{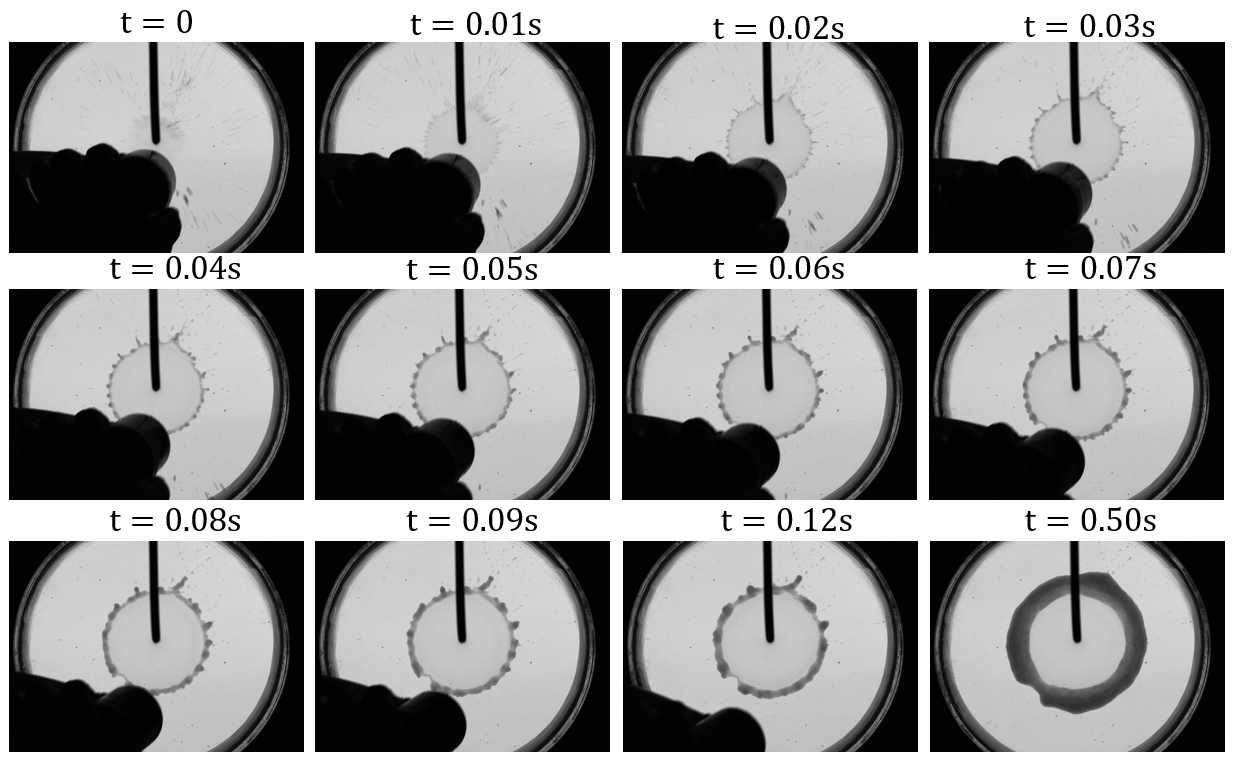}
    \caption{The transient formation of a circular hydraulic jump following the impingement of the water jet on the Perspex plate at $Q=\SI{2}{\litre\per\minute}$, \pfl{$Q^* = 97.7$}. During the initial stage, the supercritical film expands radially and the hydraulic-jump front moves outwards. From $\SI{0.08}{\second}$, the jump radius becomes stationary, while the thicker outer subcritical film continues to spread.}
    \label{Figure: jump formation}
\end{figure}

Figure~\ref{Figure: jump formation} shows a sequence of images taken with a time resolution of 0.01s from the time of impact of a water jet onto the plate .
Before starting the experiment, the liquid jet was diverted into a beaker while the flow rate was adjusted to $\SI{2}{\liter \per \minute}$, \pfl{$Q^* = 97.7$}, leaving the plate initially dry.   
At $t=0$, the beaker was rapidly withdrawn, allowing the jet to impinge on the plate and initiate the radial spreading of the film. The fluid in the jet is dyed, but is initially only visible \pfl{to the eye} after the jump, as the film is too thin between the point of impact of the jet and the jump to absorb sufficient light. The circular outer boundary of the flow moved radially outwards and had small non-axisymmetric disturbances at the front. It was observed to be thicker than the following flow and it advanced until approximately $t=\SI{0.07}{\second}$, after which it remained  stationary. At later times the flow inside this radius did not change but fluid spread beyond this in a deeper flow. By $t=\SI{0.5}{\second}$ a distinct transition between a thin inner region and a thicker outer flow is clearly established.

\begin{figure}
    \centering
    \includegraphics[width=0.8\linewidth]{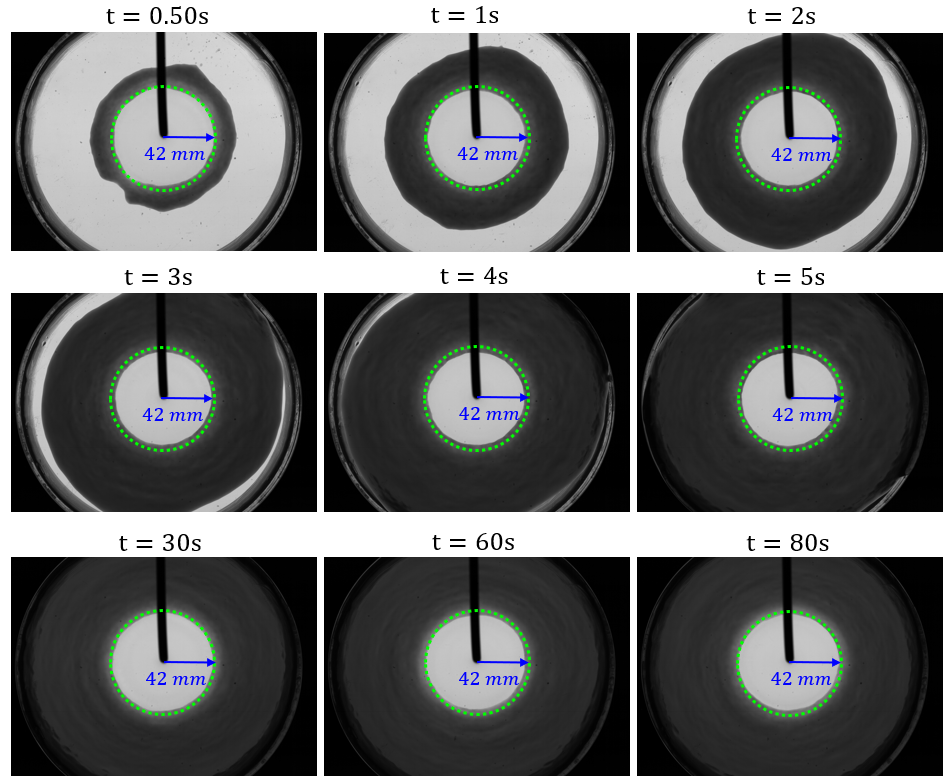}
    \caption{Later images  \pfl{of the flow shown in figure~\ref{Figure: jump formation}}. The first six images show the developing film with a moving contact line at its \pfl{outer} edge; the final three show the steady-state  \pfl{ after the front has passed into the reservoir}, leaving no contact line. A $\SI{42}{\milli\metre}$ circle is superimposed \pfl{showing} the jump radius \pfl{remains unchanged once initially established}.}
    \label{Figure: film evolution images}
\end{figure}

Figure~\ref{Figure: film evolution images} shows the subsequent evolution of the flow \pfl{shown in figure~\ref{Figure: jump formation}} as the advancing front travels radially and eventually passes beyond the edge of the plate \pfl{ at $t =\SI{5}{\second}$ when the liquid enters the reservoir}. The \pfl{location of the} jump is observed to remain at a fixed radius throughout, showing that the flow inside the jump is steady and unaffected by the flow beyond. This is characteristic of a hydraulic jump with the inner region being supercritical and unable to receive information from further downstream. 

The experiment continued uninterrupted well after the flow from the plate enters the reservoir, as shown in images at $t= 30, 60$ and $\SI{80}{\second}$ in figure \ref{Figure: film evolution images}. The experiment was designed so that, at steady state, the spreading film falls and merges seamlessly with the pool, eliminating any contact line. This experiment demonstrates that the unsteady, evolving outer film with a contact line does not influence the position of the jump (see supplementary video 1 and 2).

Figure~\ref{fig:jump-formation-film-thickness} shows the evolution of the transient film-thickness profile \pfl{\(h(r,t)\)} during the formation of the jump, as a space--time map built from every frame of the recording (\textit{a}) and as a stack of individual profiles at twelve time steps (\textit{b}). \akm{The height profiles were extracted at 10 different azimuthal angles and averaged to obtain the representative profile at each time step.} The map separates two motions: the jump, which appears as the sharp boundary between the thin supercritical film and the thick film beyond it, and the outer edge of the spreading film. Consistent with the visual observations, the jump first becomes distinguishishable at approximately \(t=\SI{0.02}{\second}\). Thereafter, both the radial extent and the height of the supercritical film increase as the jump propagates
outwards. By \(t\simeq\SI{0.08}{\second}\), the jump has reached a radius of approximately \(\SI{42}{\milli\metre}\) and becomes nearly stationary. Over the remaining frames its \pfl{location stays} at $r=$\(42.2\pm0.7\,\mathrm{mm}\).

\pfl{Beyond this time} the jump in
figure~\ref{fig:jump-formation-film-thickness}(\textit{a}) remains essentially vertical, while the film beyond the jump continues to evolve. Its outer front advances from approximately \(\SI{50}{\milli\metre}\) at \(t=\SI{0.1}{\second}\) to
\(\SI{64}{\milli\metre}\) at \(t=\SI{0.5}{\second}\), while its thickness increases from approximately \(1.4\) to \(\SI{3.0}{\milli\metre}\). Thus, over the same interval and in the
same measurement, both the extent and depth of the film beyond the jump change substantially while the jump remains fixed. This is direct \pfl{evidence} that the developing flow beyond the jump does not \pfl{affect} the jump radius \pfl{and that the flow inside the jump is steady in time}.

\begin{figure}
    \centering
    \includegraphics[width=\linewidth]{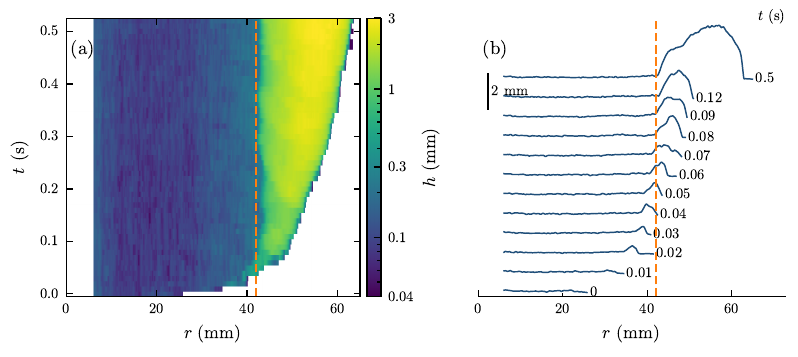}
\caption{Measured free-surface profiles $h(r)$, for the flow of
  figures~\ref{Figure: jump formation} and \ref{Figure: film evolution images}. (\textit{a}) Space--time map of
  $h(r,t)$ built from all $53$ frames of the run, $t=0$ to $0.52$~s at
  $0.01$~s, on a logarithmic colour scale, $0.04$--$3$~mm. The profiles are
  smoothed in radius; each frame is processed independently, so the map is
  additionally passed through a running median of three frames to remove the
  frame-to-frame scatter, a median rather than a mean so that the film front
  remains a step. Blank regions are inside $r=5$~mm, where the nozzle obscures
  the surface, and beyond the film front, where the film thins below the
  $0.04$~mm detection floor. (\textit{b}) Twelve of the same profiles, offset
  vertically, earliest at the bottom, labelled by time and unsmoothed; the bar
  gives the $2$~mm vertical scale. The dashed line in both panels is the
  predicted jump radius, $R_j=42$~mm.} 
 
    \label{fig:jump-formation-film-thickness}
\end{figure}

An important feature to notice about the depth profile inside the jump is that it initially decreases with radius, reaches a minimum and then increases again. The initial decrease is consistent with Bernoulli's theorem applying on the free surface, indicating that the flow is essentially inviscid \pfl{in the region near the impact point}. The increase in depth \pfl{with radius further from impact} is, therefore, associated with viscous bottom stresses. As pointed out by \cite{watson1964radial}, a viscous boundary layer grows on the plate and affects the full vertical velocity profile at a distance that scales with the Reynolds number of the jet as ${Re_j}^{\frac{1}{3}}$.

\begin{figure}
    \centering
    \includegraphics[width=\linewidth]{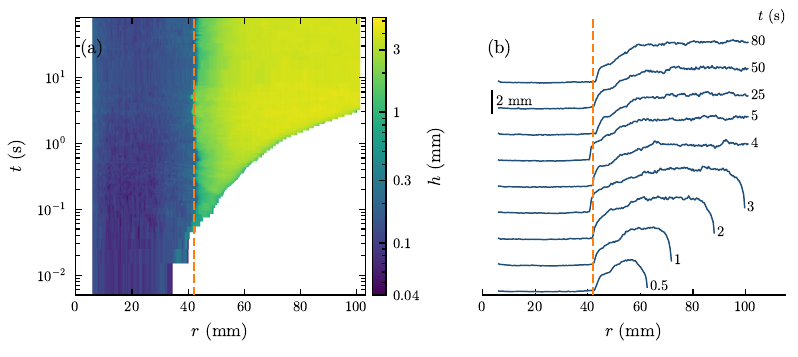}
       \caption{Later evolution of the layer height of the flow in
  figure~\ref{fig:jump-formation-film-thickness}, using the same processing. (\textit{a}) Space--time map
  of $h(r,t)$ from $119$ frames, $t=0.01$ to $80$~s, sampled evenly in
  $\log t$. The record is interrupted between $8.5$ and $17.5$~s, when the view
  was obstructed while we checked that no contact line existed near the rim and
  that the liquid fell as a continuous draining film; the profiles either side
  agree to within their scatter. (\textit{b}) Nine of the same profiles, offset
  vertically and labelled by time. The dashed line is the predicted jump
  radius, $R_j=42$~mm, which the jump holds throughout the recording. The outer
  film flows past the edge of the plate \pfl{at $t = \SI{5}{\second}$} and drains into the pool, leaving no
  contact line.}
    \label{Figure: Film evolution plot 3D and 2D}
\end{figure}

Figure~\ref{Figure: Film evolution plot 3D and 2D} extends this
observation over the full evolution of the film, from
\(t=\SI{0.01}{\second}\) to \(\SI{80}{\second}\). The outer edge of
the film continues to advance, passing approximately
\(\SI{64}{\milli\metre}\) at \(\SI{0.5}{\second}\),
\(\SI{72}{\milli\metre}\) at \(\SI{1}{\second}\), and
\(\SI{89}{\milli\metre}\) at \(\SI{2}{\second}\), before reaching the plate rim at approximately \(\SI{5}{\second}\). The film then drains continuously over the edge and into the surrounding pool, so that no solid--liquid--gas contact line remains on the plate. During this evolution the film beyond the jump also continues to deepen, approaching
its fully developed downstream thickness of approximately \(\SI{3.7}{\milli\metre}\). The record is interrupted between \(\SI{8.5}{\second}\) and \(\SI{17.5}{\second}\), while the rim was inspected to confirm that the liquid formed a continuous draining film with no contact line; the profiles on either side of the interruption agree within their scatter.

The jump, by contrast, becomes stationary very early in the evolution. From approximately \(t=\SI{0.1}{\second}\), the boundary between the thin supercritical film and the thicker film beyond it is essentially vertical in figure~\ref{Figure: Film evolution plot 3D and 2D}(\textit{a}) and
remains so for the rest of the recording. Its radius stays close to \(\SI{42}{\milli\metre}\), in agreement with the prediction from \eqref{eq:rj} shown by the orange dashed line, while the film beyond the jump continues to spread and deepen, reaches the plate rim,
and ultimately changes from a spreading film bounded by an advancing contact line to a continuously draining film with no contact line on
the plate. Thus both the downstream state and its boundary condition change substantially without displacing the jump.

This observation directly tests the proposed role of partial wetting. If the advancing contact line selected the jump radius, its continued motion and eventual disappearance would be expected to alter the jump position. No such response is observed. The result therefore provides no support for the contact-line interpretation proposed by \citet{duchesne2022circular}: the jump radius is established independently of the evolving film beyond it. We return to this question in \S\ref{sec:contact_angle}, where the wetting condition is varied directly in the numerical calculations while all other flow parameters
are held fixed.

\subsection{Quantitative measurements} \label{ssec:quant} 
\begin{figure}
    \centering
    \tikzexternaldisable
\includegraphics[width =\linewidth]{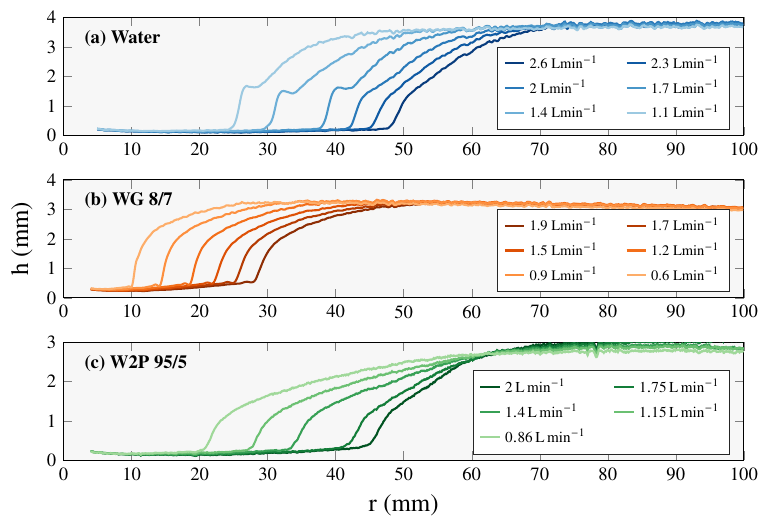}
 \caption{The measured film thickness for (a) water, (b) water + glycerol solution (WG 8/7), and (c) 2-propanol + water solutions (W2P 95/5). \pfl{Give Q* values for each case.} }
    \label{fig:film_thickness}
\end{figure}

Hereafter all measurements concern the fully developed steady state, in which the spreading film falls from the plate edge and merges with the pool, leaving no contact line. For these, each image is normalised by a background image obtained before the experiment began, one hundred frames of the steady flow are averaged, and the film thickness is then read from the calibration curve. \pfl{We present results for 3 different liquids varying by a factor of 1.6 in surface tension and 6 in viscosity, but with each case having $Q^* \gg 1$.} Figure~\ref{fig:film_thickness} (a) shows the measured film thickness of water for $Q = 1.1$--$\SI{2.6}{\litre\per\minute}$: $0.15$--$\SI{0.3}{\milli\metre}$ upstream of the jump, varying weakly with flow rate, and $\approx\SI{3.7}{\milli\metre}$ beyond it. \akm{Figure~\ref{fig:film_thickness}(b) presents the corresponding measurements for WG8/7 over $Q=\SIrange{0.6}{1.9}{\litre\per\minute}$. In this case, the supercritical film thickness varies from approximately $\SI{0.24}{\milli\metre}$ to $\SI{0.38}{\milli\metre}$, while the post-jump thickness is approximately $\SI{3.2}{\milli\metre}$. Figure~\ref{fig:film_thickness}(c) shows the film profiles for W2P 95/5 over $Q=\SIrange{0.86}{2.0}{\litre\per\minute}$. The supercritical film thickness ranges from approximately $\SI{0.13}{\milli\metre}$ to $\SI{0.35}{\milli\metre}$, and the downstream film thickness is approximately $\SI{2.9}{\milli\metre}$.} In all cases the super-critical film is slowly varying, and the jump marks a sharp transition from supercritical to sub-critical film.  
 
\pfl{While the jump location is clearly increases with flowrate in every case, the dependence of the film thickness and the jump location on the liquid properties is not obvious from these profiles. The film inside the jump is significantly thicker for the most viscous flow of WG8/7, but the location of the jump is not obviously set by surface tension alone. For example, for $Q = \SI{2}{\litre\per\minute}$ the jump positions for water and WP295/5, whose surface tensions vary by a factor of 1.6, are essentially the same at 46mm. On the other hand, the jump position is closer to the axis for WG8/7 compared with water, although they both have similar values of $\gamma$. For this reason, we look at the values of the governing non-dimensional parameters in the next section.  }

\begingroup

\newcommand{\colw}{0.29\textwidth}     
\newcommand{\rowlabw}{0.06\textwidth}  

\newcommand{\colhead}[1]{%
    \makebox[\colw][c]{\textbf{\Large #1}}%
}

\newcommand{\rowhead}[1]{%
    \makebox[\rowlabw][c]{%
        \rotatebox{90}{\textbf{\large #1}}%
    }%
}

\newcommand{\fitgraphic}[1]{%
    \includegraphics[width=4.5cm]{#1}%
}

\captionsetup[subfigure]{justification=centering}

\begin{figure}
    \centering

    \makebox[\rowlabw]{}%
    \colhead{\hspace{1.2cm}{$Bo$}}\hfill
    \colhead{\hspace{1.2cm}{$Fr^2$}}\hfill
    \colhead{\hspace{1.2cm}{$We$}}\\[0.6em]

    \rowhead{\hspace{1.8cm}Water}%
    \begin{subfigure}[t]{\colw}
        \centering
        \fitgraphic{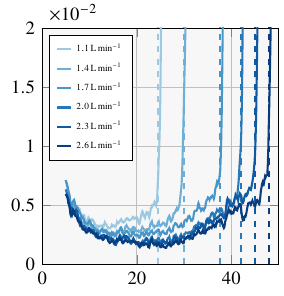}
        \caption{}
    \end{subfigure}\hfill
    \begin{subfigure}[t]{\colw}
        \centering
        \fitgraphic{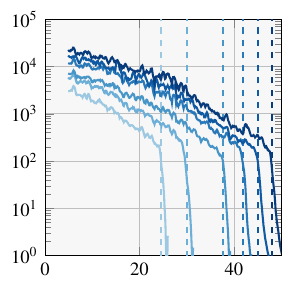}
        \caption{}
    \end{subfigure}\hfill
    \begin{subfigure}[t]{\colw}
        \centering
        \fitgraphic{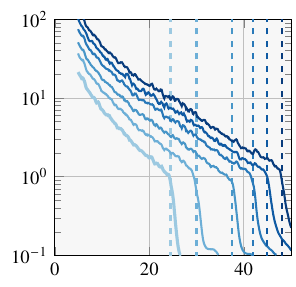}
        \caption{}
    \end{subfigure}\\[0.9em]

    \rowhead{\hspace{1.8cm}WG 8/7}%
    \begin{subfigure}[t]{\colw}
        \centering
        \fitgraphic{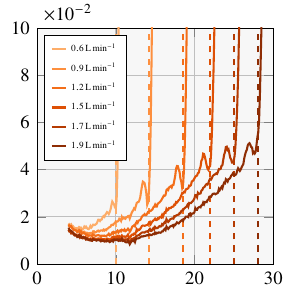}
        \caption{}
    \end{subfigure}\hfill
    \begin{subfigure}[t]{\colw}
        \centering
        \fitgraphic{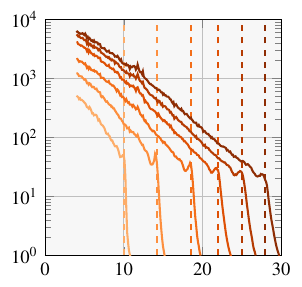}
        \caption{}
    \end{subfigure}\hfill
    \begin{subfigure}[t]{\colw}
        \centering
        \fitgraphic{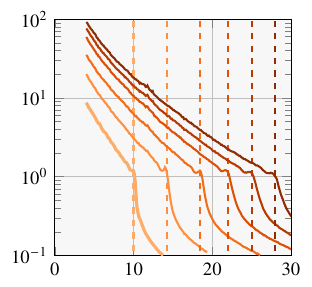}
        \caption{}
    \end{subfigure}\\[0.9em]

    \rowhead{\hspace{1.8cm}W2P 95/5}%
    \begin{subfigure}[t]{\colw}
        \centering
        \fitgraphic{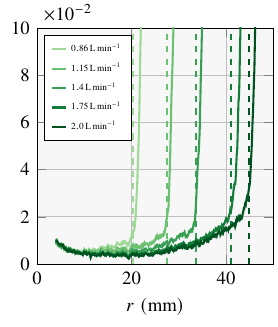}
        \caption{}
    \end{subfigure}\hfill
    \begin{subfigure}[t]{\colw}
        \centering
        \fitgraphic{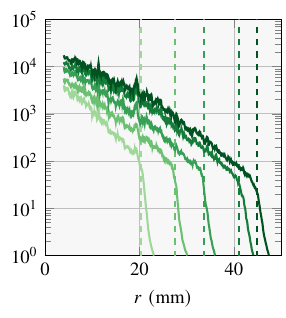}
        \caption{}
    \end{subfigure}\hfill
    \begin{subfigure}[t]{\colw}
        \centering
        \fitgraphic{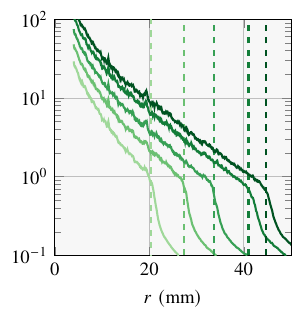}
        \caption{}
    \end{subfigure}\\[0.4em]

    \caption{
        Local Bond number ($Bo$), squared Froude number ($Fr^2$),
        and Weber number ($We$) in the supercritical flow for water,
        WG~8/7, and W2P~95/5, computed using the depth-averaged velocity.
        Coloured curves represent different flow rates, consistent with
        figure~\ref{fig:film_thickness}(a), while the dashed vertical
        lines indicate the corresponding hydraulic-jump locations.
    }
    \label{fig:water_WG_W2P_Bo_Fr_We}

\end{figure}

\endgroup

%
\subsection{Local Bond, Froude and Weber numbers}
\label{sec:Bo_Fr_We}

The film-thickness measurements give the Bond number directly through \eqref{eq:Bo}, and the depth-averaged velocity through continuity,
\(Q=2\pi r h U\). The local Froude and Weber numbers can, therefore, be evaluated as functions of radius from the measured \(h(r)\) alone, along with
 the imposed flow rate and the fluid properties.

Figure~\ref{fig:water_WG_W2P_Bo_Fr_We} shows the three dimensionless groups 
\pfl{as functions of radius} for every liquid and flow rate examined, with the measured jump radii marked by dashed lines. Throughout the supercritical film, \(Bo=O(10^{-3})\)--\(O(10^{-1})\), so gravity is weak relative to surface tension \pfl{as expected since $Q^* \gg 1$ in each case}. As expected, since $Uh$ is a decreasing function of radius, both $We$ and $Fr^2$, although initially large, decrease away from the point of impact of the jet. At the measured jump radii, the values of \(Fr^2\) remain large, \(Fr^2=O(10)\)--\(O(10^2)\), and vary substantially
between fluids and flow rates. There is therefore no common Froude number associated with the jump. The corresponding Weber numbers show a strikingly different result: despite the variation in fluid properties and flow rate, the Weber number at the jump is of order unity in every case. 

This distinction is the central experimental observation. The jump locations do not correspond to a common gravitational critical state, whereas they do correspond to a common capillary condition,
\(We=O(1)\). Moreover, \eqref{eq:Bo} gives \(Fr^2=\frac{We}{Bo},\) so that the coexistence of \(We=O(1)\) and \(Bo\ll1\) necessarily
places the jump in a regime with \(Fr^2\gg1\), as observed. The jump therefore occurs while the film remains strongly supercritical with respect to gravity waves. 
\pfl{We emphasise that the experiments have been chosen to be within the regime defined by $Q^* \gg 1$. Under different conditions, such as much higher jet flow rates, we expect lower values of the $Fr$ and for gravity to play a role. These cases will be discussed in a future paper. }
\rkb{
\subsection{Jump radius}
\label{subsec:jump_location}
 
 \pfl{In the discussion above we have provided a predicted radius $R_j$ of the jump. It is derived as follows.}  The experimental observations show that at the jump \(Bo\ll1\),
\(Fr^{2}\gg1\) and \(We=O(1)\). \pfl{Consequently, the} supercritical film is \pfl{thin and is governed }
by the viscous shallow-water equation
\citep{tani1949water,bohr1993shallow,wang2021effects}. \pfl{Since, for $r \leq R_j$}, \(Fr^{2}\gg1\) the
hydrostatic pressure gradient is negligible and the leading-order
equation is, as shown in Appendix~\ref{app:film_scaling}, 
\begin{equation}
    u\frac{\partial u}{\partial r}
    +w\frac{\partial u}{\partial z}
    =\nu\frac{\partial^{2}u}{\partial z^{2}}.
\label{eq:parameter_free}
\end{equation}
The boundary conditions are no slip at the wall,
\(u=0\) at \(z=0\), and zero shear stress at the free surface,
\(\partial u/\partial z=0\) at \(z=h\). Depth integration of
\eqref{eq:parameter_free} \pfl{and continuity} give
\begin{equation}
    \frac{\mathrm{d}}{\mathrm{d}r}
    \left(r\int_{0}^{h}u^{2}\,\mathrm{d}z\right)
    =-\nu r\left.\frac{\partial u}{\partial z}\right|_{z=0}.
\label{eq:momentum_integral}
\end{equation}
which can be closed with a boundary-layer velocity field \(u=U f(\eta)\), where \(U\) is the depth-average velocity, \(\eta=z/h\) and \(f(\eta)\) is a velocity shape function.
For the gravity-free radial film,
\citet{watson1964radial} obtained the corresponding exact similarity
solution
\begin{equation}
    h^{2}\frac{\mathrm{d}U}{\mathrm{d}r}
    =-\frac{3}{2}c^{2}I_{1}\nu,
\label{eq:watson}
\end{equation}
where the shape factors
\(I_{1}=\int_{0}^{1}f(\eta)\,\mathrm{d}\eta=0.615\) and
\(c=f'(0)=1.402\).
Solving \eqref{eq:watson} or \eqref{eq:momentum_integral} together
with volume-flux conservation,
\begin{equation}
    Urh=q=\frac{Q}{2\pi},
\label{eq:volumeflux}
\end{equation}
and applying the capillary-critical condition
\begin{equation}
    We=\frac{\rho U^{2}h}{\gamma}=1,
\label{eq:We_closure}
\end{equation}
gives, far from the jet (\(r^{3}\gg l^{3}\), with \(l\) the constant
of integration of \eqref{eq:watson}), the jump radius
\begin{equation}
    \frac{R_{j}}{R_{\gamma}}
    =\left(\frac{3I_{2}}{I_{1}}\right)^{1/4}
    \approx1.13,
\label{eq:rc}
\end{equation}
where \(
    I_{2}\equiv\frac{1}{c}\int_{0}^{1}f^{2}\,\mathrm{d}\eta=0.339\). In terms of the imposed flow rate
\(Q\),
\begin{equation}
    R_{j}
    =\left(\frac{3I_{2}}{8\pi^{3}I_{1}}\right)^{1/4}
    \left(\frac{\rho Q^{3}}{\nu\gamma}\right)^{1/4}
    \approx0.286
    \left(\frac{\rho Q^{3}}{\nu\gamma}\right)^{1/4}.
\label{eq:rj}
\end{equation}
The prefactor is also only weakly sensitive to the velocity
shape function $f(\eta)$. Solving \eqref{eq:momentum_integral} with the parabolic
shape function gives \(0.264\), \(7.7\,\%\) below the Watson value
\(0.286\), while the cubic shape function of \citet{wang2019role}
gives \(0.281\), only \(1.7\,\%\) below the Watson value. Using instead a Weber number based on the momentum flux,
\(\rho\int_{0}^{h}u^{2}\,\mathrm{d}z/\gamma\),
\citet{bhagat2018origin} obtained
\(R_{j}=0.277(\rho Q^{3}/\nu\gamma)^{1/4}\), \(3\,\%\) below the value in \eqref{eq:rj}. In the plots, we used \eqref{eq:rj} to predict the jump radii.
}

\section{Numerical Investigation}
\label{sec:numerics}

We also investigate the flow numerically. The air–liquid interface is resolved using the \texttt{interFoam} solver in OpenFOAM, which employs a finite-volume formulation for transient, incompressible and isothermal flows of two immiscible Newtonian fluids. The interface is captured using the volume-of-fluid (VOF) method, in which a phase-fraction field $\alpha$ is advected throughout the computational domain. The \texttt{interFoam} solver has been extensively assessed against analytical solutions and experimental measurements and has been shown to provide reliable predictions for inertia-dominated, free-surface and multiphase flows \citep{deshpande2013evaluating,bayon2015numerical,bayon2016performance,mukha2022large,askarizadeh2019role}. The VOF method permits large interface deformations without requiring explicit tracking of the interface topology. To reduce numerical diffusion of the phase-fraction field, \texttt{interFoam} introduces an anti-diffusive compression flux directed approximately normal to the interface. This interface-compression term imposes an artificial compression velocity within the interfacial region and thereby maintains a sharp transition between the liquid and gas phases.

\subsection{Numerical Formulation}

We applied the Navier-Stokes equations within the volume of fluid (VOF) framework, as described by \cite{hirt1981volume}. The governing equations are the incompressibility condition,
\begin{equation}
    \frac{\partial u_i}{\partial x_i}=0,
    \label{eq:continuity_interfoam}
\end{equation}
and the momentum equation,
\begin{equation}
    \frac{\partial (\rho u_i)}{\partial t}
    +
    \frac{\partial (\rho u_i u_j)}{\partial x_j}
    =
    -\frac{\partial p}{\partial x_i}
    +
    \frac{\partial}{\partial x_j}
    \left[
        \mu
        \left(
            \frac{\partial u_i}{\partial x_j}
            +
            \frac{\partial u_j}{\partial x_i}
        \right)
    \right]
    +
    \rho g_i
    +
    F_i^{\gamma},
    \label{eq:momentum_interfoam}
\end{equation}
where $u_i$ denotes the velocity components, $p$ is the pressure, $\rho$ is
the local mixture density, $\mu$ is the local dynamic viscosity, $g_i$ denotes
the gravitational acceleration, $F_i^{\gamma}$ is the volumetric
surface-tension force, \pfl{and the summation convention applies}. The density and viscosity are evaluated from the phase fraction according to
\begin{equation}
    \rho
    =
    \alpha \rho_{\ell}
    +
    (1-\alpha)\rho_{g},
    \qquad
    \mu
    =
    \alpha \mu_{\ell}
    +
    (1-\alpha)\mu_{g},
    \label{eq:mixture_properties}
\end{equation}
where the subscripts $\ell$ and $g$ denote the liquid and gas phases,
respectively. In the OpenFOAM implementation, the momentum equation is
formulated using the hydrostatically modified pressure, $p_{\mathrm{rgh}} = p-\rho\,\boldsymbol{g}\boldsymbol{\cdot}\boldsymbol{x}$, which separates the hydrostatic contribution from the pressure field. The phase fraction is transported using
\begin{equation}
    \frac{\partial \alpha}{\partial t}
    +
    \frac{\partial (\alpha u_i)}{\partial x_i}
    +
    \frac{\partial}{\partial x_i}
    \left[
        \alpha(1-\alpha)u_{c,i}
    \right]
    =0,
    \label{eq:alpha_transport}
\end{equation}
where, \pfl{as mentioned above,} $u_{c,i}$ is an artificial compression velocity directed approximately
normal to the interface. The third term in
\eqref{eq:alpha_transport} compresses the transition region between the two
phases and thereby limits numerical diffusion of the phase-fraction field. The
factor $\alpha(1-\alpha)$ confines this compression to interfacial cells and
causes the term to vanish in regions occupied entirely by either fluid. Surface tension is incorporated using the continuum-surface-force formulation
of \citet{brackbill1992continuum}, in which the interfacial force is distributed
over the finite-thickness numerical interface \pfl{in the form}
\begin{equation}
    \boldsymbol{F}^{\gamma}
    =
    \gamma \kappa \boldsymbol{\nabla}\alpha,
    \label{eq:csf_force}
\end{equation}
where $\gamma$ is the surface-tension coefficient and $\kappa$ is the
interface curvature. The curvature is obtained from
\begin{equation}
    \kappa
    =
    -\boldsymbol{\nabla}\boldsymbol{\cdot}\widehat{\boldsymbol{n}},
    \qquad
    \widehat{\boldsymbol{n}}
    =
    \frac{\boldsymbol{\nabla}\alpha}
    {\left|\boldsymbol{\nabla}\alpha\right|+\delta_n},
    \label{eq:interface_curvature}
\end{equation}
where $\widehat{\boldsymbol{n}}$ is the unit normal to the interface and
$\delta_n$ is a small regularisation parameter introduced to avoid division
by zero away from the interface. The sign of
\eqref{eq:csf_force} depends on the convention adopted for the interface
normal and curvature; the definitions in
\eqref{eq:interface_curvature} are used consistently here. We discretise the computational domain in OpenFOAM using the interFoam solver where the flow interface is represented by mixed cells containing both fluids. In the context of our simulation, $\alpha = 1$ represents the liquid phase while $\alpha = 0$ represents the air phase and at the interface $\alpha =0.5$. These equations collectively describe the conservation of mass, momentum, and phase distribution in the multiphase flow.

\subsection{\label{sec:sim domain}Simulation Domain }

\begin{figure}
\centering
\includegraphics[width=0.8\linewidth]{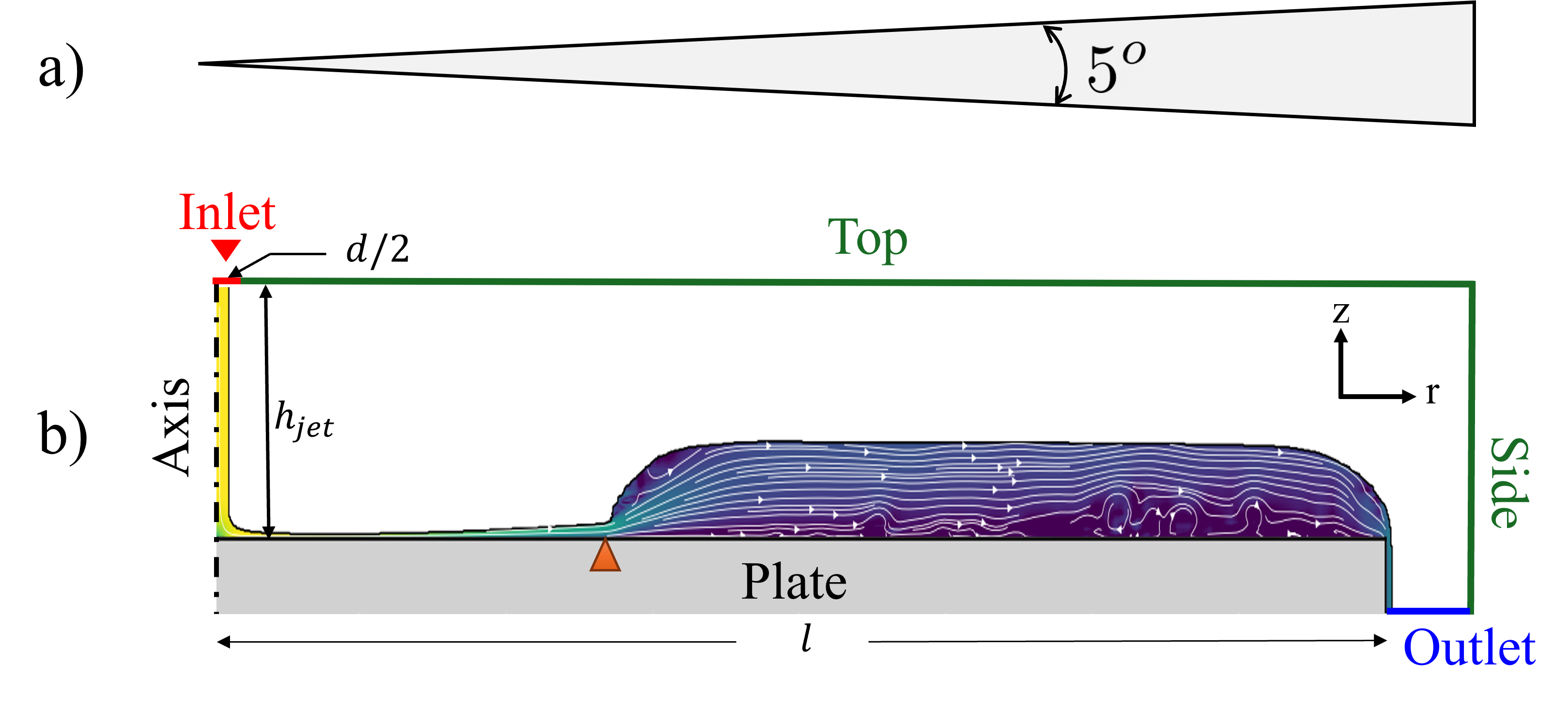}
\caption{The figure shows the schematic of the simulation domain for 2D axisymmetric flow. (a) shows the top view of the domain a wedge, \pfl{centered beneath the jet axis,} of angle $5^o$ . (b) shows the front view with the different boundaries. Here, $d$ denotes the jet diameter, $l$ the plate length, and $h_{\text{jet}}$ the inlet height from the plate. The jump is denoted by the orange triangle at $r = R_j$. The white curves represent the streamlines. Note that, while streamlines appear to end this is an artifact of the contouring used. }
\label{Figure: simulation domain schematic}
\end{figure}

To simulate the circular hydraulic jump, we employ a simple and computationally efficient two-dimensional axisymmetric model, as shown in figure~\ref{Figure: simulation domain schematic}. The computational domain consists of a wedge with an angular extent of $5^\circ$, with its apex centred on the axis of symmetry and spanning equally either side of $x=0$ i.e., the central azimuthal plane. In this representation, the $x$--direction corresponds to the radial direction of the flow, whereas the $z$--direction represents the vertical and hence the liquid-film thickness. The jet diameter $d$ is set to $\SI{2}{\milli\metre}$ in all simulations, except for the cases used for direct experimental validation, for which $d$ is matched to the corresponding experimental value. The jet height $h_{\mathrm{jet}}$ is fixed at $\SI{10}{\milli\metre}$ for all cases. The radial extent of the plate $l$ is chosen to be approximately three times the expected jump radius $R$ in order to minimise the influence of the downstream boundary on the jump region. The height of the outlet patch is set to $\SI{3}{\milli\metre}$, consistent with the downstream liquid depth used in the experiments. \rkb{In addition, simulations are performed under the exact experimental conditions, with the jet diameter, plate radius and fluid properties matched to those of the experiments, enabling a direct one-to-one comparison between the simulated and measured film-thickness profiles.}
\par
The boundary conditions applied to the computational domain are summarised in table~\ref{tab:bcs}. A uniform velocity and fixed-pressure condition are prescribed at the jet inlet, while a no-slip condition is imposed on the plate. Pressure inlet--outlet conditions are applied at the remaining open boundaries. Wedge boundary conditions are imposed on the two azimuthal faces, thereby enforcing axisymmetry by constraining the solution to remain invariant in the circumferential direction. Similar boundary conditions were employed by \citet{askarizadeh2019role} in their numerical investigation of free-surface radial flows.

\begin{table}
\caption{\label{tab:bcs} Boundary conditions for the computational domain.}
\centering
\small
\renewcommand{\arraystretch}{1.6}
\setlength{\tabcolsep}{5pt}

\begin{tabular}{lccc}
\hline
\textbf{Boundary} & \textbf{$p$} & \textbf{$\mathbf{u}$} & \textbf{$\alpha$} \\
\hline

Axis
&
$\mathbf{n}\cdot\nabla p = 0$
&
$u_n=0,\quad \mathbf{n}\cdot\nabla\mathbf{u}_t=0$
&
$\mathbf{n}\cdot\nabla\alpha=0$
\\

Inlet
&
$\mathbf{n}\cdot\nabla p = 0$
&
$\mathbf{u}=-U_{\mathrm{jet}}\mathbf{e}_z$
&
$\alpha=1$
\\

Plate
&
$\mathbf{n}\cdot\nabla p = 0$
&
$\mathbf{u}=0$
&
$\mathbf{n}\cdot\nabla\alpha=0$
\\

\multirow{2}{*}{Top, Side \& Outlet}
&
$p=p_0$
&
$\mathbf{n}\cdot\nabla\mathbf{u}=0$
&
$\mathbf{n}\cdot\nabla\alpha=0$
\\[-1mm]

& 
$p=p_0-\dfrac{1}{2}|\mathbf{u}|^2$
&
$\mathbf{u}_t=0,\quad u_n\text{ determined by flux}$
&
$\alpha=0$
\\

Wedge faces
&
$\mathbf{n}\cdot\nabla p=0$
&
$u_n=0,\quad \mathbf{n}\cdot\nabla\mathbf{u}_t=0$
&
$\mathbf{n}\cdot\nabla\alpha=0$
\\

\hline
\end{tabular}

\vspace{1mm}
\parbox{0.95\linewidth}{\footnotesize
Here, $\mathbf{n}$ and $\mathbf{t}$ denote the outward normal and tangential
directions to a boundary, respectively. For the Top, Side and Outlet boundaries,
the first and second rows correspond to outflow
($\mathbf{u}\cdot\mathbf{n}>0$) and inflow
($\mathbf{u}\cdot\mathbf{n}<0$), respectively.
The phase fraction $\alpha=1$ denotes the liquid phase, while
$\alpha=0$ denotes the gas phase.}
\end{table}

\subsection{Numerical considerations and simulation protocol}
\label{sec:num_protocol}
All fluids used in the experiments were simulated, while additional fluids considered only in the numerical simulations are listed in Table~\ref{tab:fluid properties for simulations}. \pfl{The model fluids (MF) were chosen with the same density and viscosity as water, but with different values of surface tension.}
 \begin{table}
\caption{Fluid properties and flow-rate ranges used for additional simulations}
\label{tab:fluid properties for simulations}
\centering
\begin{tabular}{ccccccc}
\hline
\textbf{Fluid} & \textbf{$\rho$ ($\SI{}{\kilogram \meter^{-3}}$)} & \textbf{$\nu$ ($\SI{}{\meter^2 \second^{-1}}$)} & \textbf{$\gamma$ ($\SI{}{\newton \meter^{-1}}$)}& Q $(\SI{}{\liter \per \minute})$ & \textbf{$Q^*$} & \makecell{Reference for \\ fluid properties} \\
\hline
Water+SDBS & $1000$ & $1 \times 10^{-6}$ & 0.038 & 0.6 - 1.3 & 42.7 - 92.5 &  \cite{bhagat2018origin} \\ \\

MF1 & $1000$ & $1 \times 10^{-6}$ & 0.01 & 1 & 3.8 &  -  - \\ \\
MF2 & $1000$ & $1 \times 10^{-6}$ & 0.03 & 1 & 34.6 &  -  - \\ \\
MF3 & $1000$ & $1 \times 10^{-6}$ & 0.05 & 1 & 96 &  -  - \\ 
\end{tabular}

\vspace{1mm}

\raggedright
\footnotesize
$^{*}$MF = Model fluids.
\end{table}
The principal numerical difficulty in these calculations arises from the
treatment of the surface-tension-dominated interface rather than from the
hydraulic jump itself. In volume-of-fluid simulations with
continuum-surface-force coupling \citep{brackbill1992continuum},
discretisation errors in the computed curvature can generate parasitic
currents. In the viscous limit their characteristic magnitude scales as
\(u_{\rm spur}\sim C\,\frac{\gamma}{\mu},\) with benchmark calculations giving \(C=O(10^{-2})\)
\citep{lafaurie1994modelling,harvie2006analysis,popinet2018numerical}.
This estimate is used here as a measure of numerical susceptibility rather than as a prediction of the parasitic velocity in the present flow, since
the actual magnitude also depends on the local interface geometry and on
the balance between viscous, inertial and capillary effects. Surface tension
also imposes a capillary-wave time-step restriction,
\(\Delta t \lesssim\left(\frac{\rho\,\Delta x^{3}}{2\pi\gamma}\right)^{1/2},\)
which becomes increasingly restrictive as the mesh is refined or the
surface tension is increased \citep{brackbill1992continuum,denner2015numerical}.

The capillary-velocity scale nevertheless provides a useful comparison
between the liquids considered here. For the model fluids and WG~8/7,
\(\gamma/\mu\approx10\)--\(50\,\mathrm{m\,s^{-1}}\), corresponding to a
benchmark parasitic-current scale of at most approximately
\(0.5\,\mathrm{m\,s^{-1}}\) for \(C=10^{-2}\). For water,
\(\gamma/\mu\approx72\,\mathrm{m\,s^{-1}}\), giving a corresponding scale
of approximately \(0.7\,\mathrm{m\,s^{-1}}\). Water is therefore,
a priori, the most demanding fluid numerically. These values should not be
interpreted as the parasitic velocities actually present throughout the
computed films; the comparison is local, and the numerical susceptibility
differs markedly between regions of the flow.

In the supercritical film the velocity is of order \(\SI{1}{\meter\per\second}\) and the free surface is nearly flat. The source of curvature error is therefore weak, while the physical flow is strongly inertia dominated---the regime in which \texttt{interFoam} has been found to perform reliably \citep{deshpande2013evaluating}. Beyond the jump the velocity falls to
\(O(10^{-2}) \SI{}{\meter\per\second}\). \rkb{In this outer film gravity is no
longer negligible: its depth is comparable to the capillary length,
so that gravitational and capillary effects are both important in
determining the large-scale free-surface shape. This does not remove
the sensitivity to the numerical treatment of surface tension, since
parasitic currents arise from the discretisation of interface
curvature wherever that curvature is appreciable. The jump face and,
particularly, the draining film at the plate rim contain the strongest
interface curvature in the domain. Numerical effects associated with
the surface-tension discretisation are consequently expected to be
most important in the slowly moving outer film and in the strongly
curved regions at the jump face and plate rim. The location of the
jump, by contrast, is selected by the upstream supercritical film,
where the numerical conditions are most favourable.} 

All calculations use the mesh, solver settings and convergence tolerances described in appendix~\ref{sec:mesh}, together with adaptive time stepping bounded by flow and interface Courant numbers of \(0.5\). Unless stated otherwise, the imposed contact angle between the liquid and the plate is set to \(30^\circ\). The transient experiment--simulation comparison of \S\ref{sec:Transient state} uses the reported advancing contact angle, with the reported static value used as a check; and the water calculation requires \pfl{ a new} treatment described below. Each calculation is continued until the jump position and the resolved
free-surface and velocity fields cease to exhibit systematic temporal evolution. Under this protocol WG~8/7 and the model fluids attain steady
states, with WG~8/7 doing so for every contact angle examined in \S\ref{sec:contact_angle}.

Water is the exception. With a finite imposed contact angle, the film initially spreads to the plate rim and begins to drain, but the draining
sheet subsequently breaks at the rim and reforms. This process repeats,
producing persistent oscillations in the downstream flow and preventing
the calculation from reaching a steady state. A similar behaviour was reported by \cite{fernandez2019origin}. Significantly, the
difficulty remains confined to the subcritical film and the plate edge:
despite these oscillations, the upstream supercritical film and the jump
radius remain unchanged.

The steady water calculation was therefore performed in two stages. The
film was first allowed to spread to the plate edge with \(\theta=30^\circ\). Once drainage had
begun, the imposed wall contact angle was set to \(\theta=0^\circ\).
This maintains a continuous wetting film over the rim, prevents the
repeated breakup of the draining sheet, and allows the entire calculation
to converge to a steady state.

This treatment is consistent with the corresponding experimental
configuration in which the fully developed film flows continuously over the plate edge
and merges with the surrounding liquid, so that no solid--liquid--gas contact line remains on the plate. The repeated breakup obtained numerically for water at finite imposed contact angle is therefore absent from the experimental steady state. More importantly, the jump is insensitive to the downstream wetting treatment. The calculations of \S\ref{sec:contact_angle} vary the imposed contact angle from \(0^\circ\) to \(90^\circ\), producing large changes in the spreading and structure of the subcritical film but no measurable change in the steady
jump radius. The plate rim also lies far downstream of the jump. The water
jump radius obtained with \(\theta=0^\circ\) agrees with the prediction
\eqref{eq:rj} and with the measured profiles.


\subsection{Comparison with experiments}
\subsubsection{Transient state} 
\label{sec:Transient state}
\begin{figure}
    \centering
    \includegraphics[width=\linewidth]{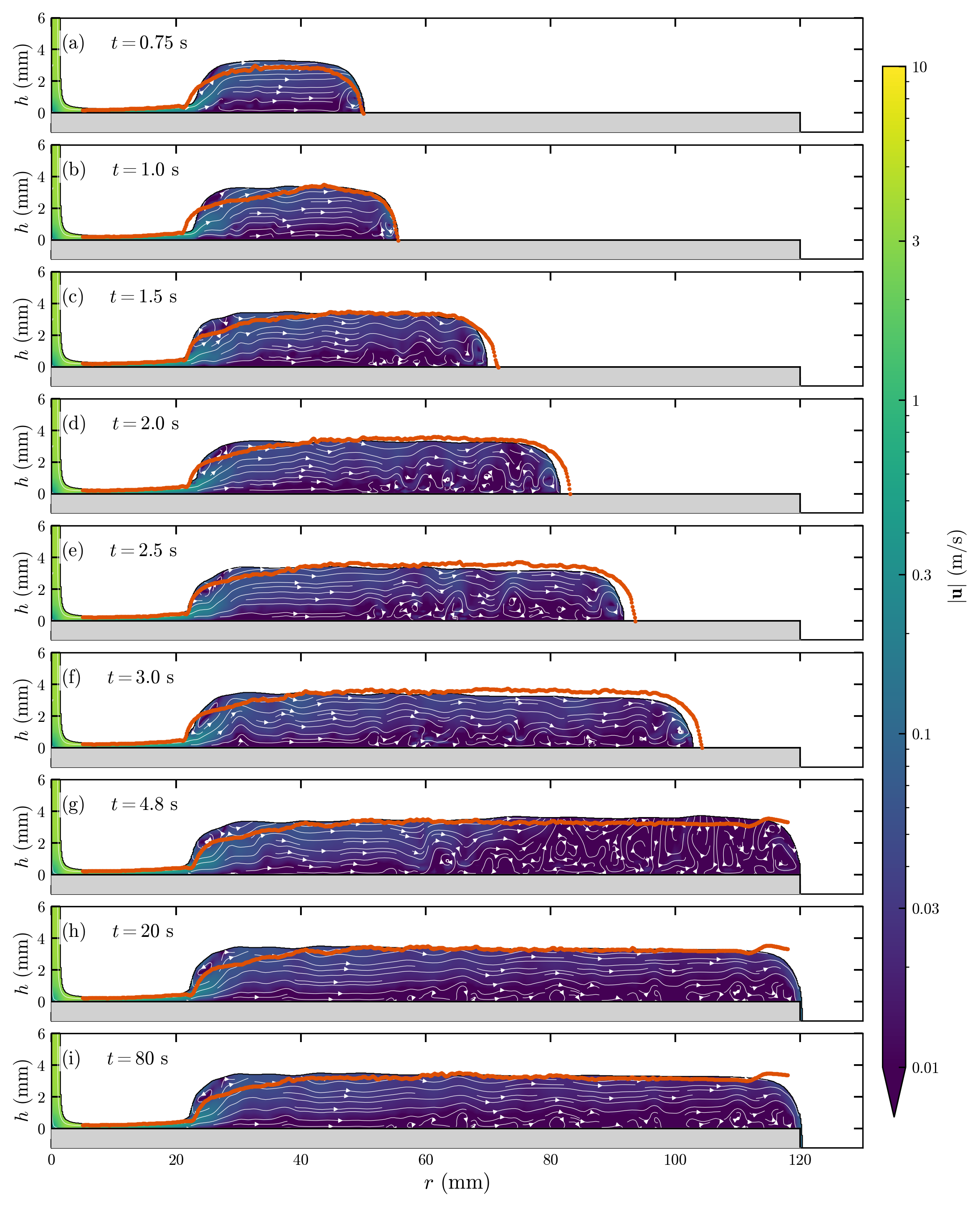}
\caption{Simulated flow field and the measured film thickness for WG~8/7 at \(Q=\SI{1.5}{\liter\per\minute}\), $Q^* = 14.8,$ produced by a
\(\SI{3}{\milli\metre}\)-diameter jet impinging on a \(\SI{240}{\milli\metre}\)-diameter Perspex plate. The simulation shown uses an imposed contact angle of \(\theta=90^\circ\), corresponding to the reported advancing contact angle for WG~8/7.  Panels (a)--(i) show \(t=0.75\), \(1\), \(1.5\), \(2\), \(2.5\), \(3\), \(4.8\), \(20\) and \(80\,\mathrm{s}\), respectively.
Colour denotes the velocity magnitude \(|\mathbf{u}|\) in log scale, white curves are
streamlines,  the orange curves show the free surface \pfl{measured in the experiments}, and the grey block represents the plate. The vertical scale is exaggerated by a factor of approximately 2. By \(t\approx4.8\,\mathrm{s}\) the outer film reaches the plate rim and drains over it, so that no
solid--liquid--gas contact line remains on the plate thereafter.}
    \label{Figure: Film evolution Expt vs Sim}
\end{figure}

Figure~\ref{Figure: Film evolution Expt vs Sim} compares the simulated and measured free-surface profiles for WG~8/7 at \(Q=\SI{1.5}{\liter\per\minute}\), \pfl{$Q^* = 14.8$}, produced by a
\(\SI{3}{\milli\metre}\)-diameter jet impinging on a \(\SI{240}{\milli\metre}\)-diameter Perspex plate. The simulation imposes a contact angle of \(\theta=90^\circ\), corresponding to the reported advancing contact angle for WG~8/7 \citep{quetzeri2019role}. The calculation was also repeated with \(\theta=75^\circ\), corresponding to the reported static contact angle \citep{sharp2011resonant,zhang2021characteristics}, with no appreciable change in the jump location or the agreement with experiment. Nine instants are shown between \(t=\SI{0.75}{\second}\) and
\(\SI{80}{\second}\). During the early evolution the outer film continues to spread, reaching the plate rim at approximately \(t=\SI{4.8}{\second}\). It then drains over the rim, so that no solid--liquid--gas contact line remains on the plate. The simulation was continued beyond \(\SI{80}{\second}\) in this freely draining state.

The numerical calculation reproduces the measured evolution closely. Within the supercritical film the simulated and measured free surfaces
agree to within \(\SI{0.08}{\milli\metre}\) r.m.s., while the larger discrepancies are confined principally to the strongly curved jump and outer-film regions. Throughout the evolution the jump radius remains at approximately \(\SI{22}{\milli\metre}\), to within \(\SI{0.8}{\milli\metre}\), while the advancing front moves from approximately \(52\) to \(\SI{120}{\milli\metre}\). Thus the jump remains fixed while the downstream film spreads, reaches the rim and loses its contact line.

\rkb{Note that the experimental measurements correspond to instantaneous images, and the subcritical film spreads asymmetrically owing to
surface roughness and other plate irregularities. The simulations, performed on a \(5^\circ\) axisymmetric wedge, cannot reproduce this
random azimuthal variation, which accounts for the small mismatch near the plate rim. The imposed fixed contact angle may also contribute to
this discrepancy. Optical lensing has additionally been identified near the strongly curved jump face, where it locally distorts the apparent film thickness. The affected points are confined to the immediate vicinity of the jump face and do not influence the supercritical film
or the measured jump radius. Despite these local differences, the simulations remain in excellent agreement with the experiments.  }

\subsubsection{Steady state}
\label{sec: steady state}
\begin{figure}
    \centering
\includegraphics[width = \linewidth]{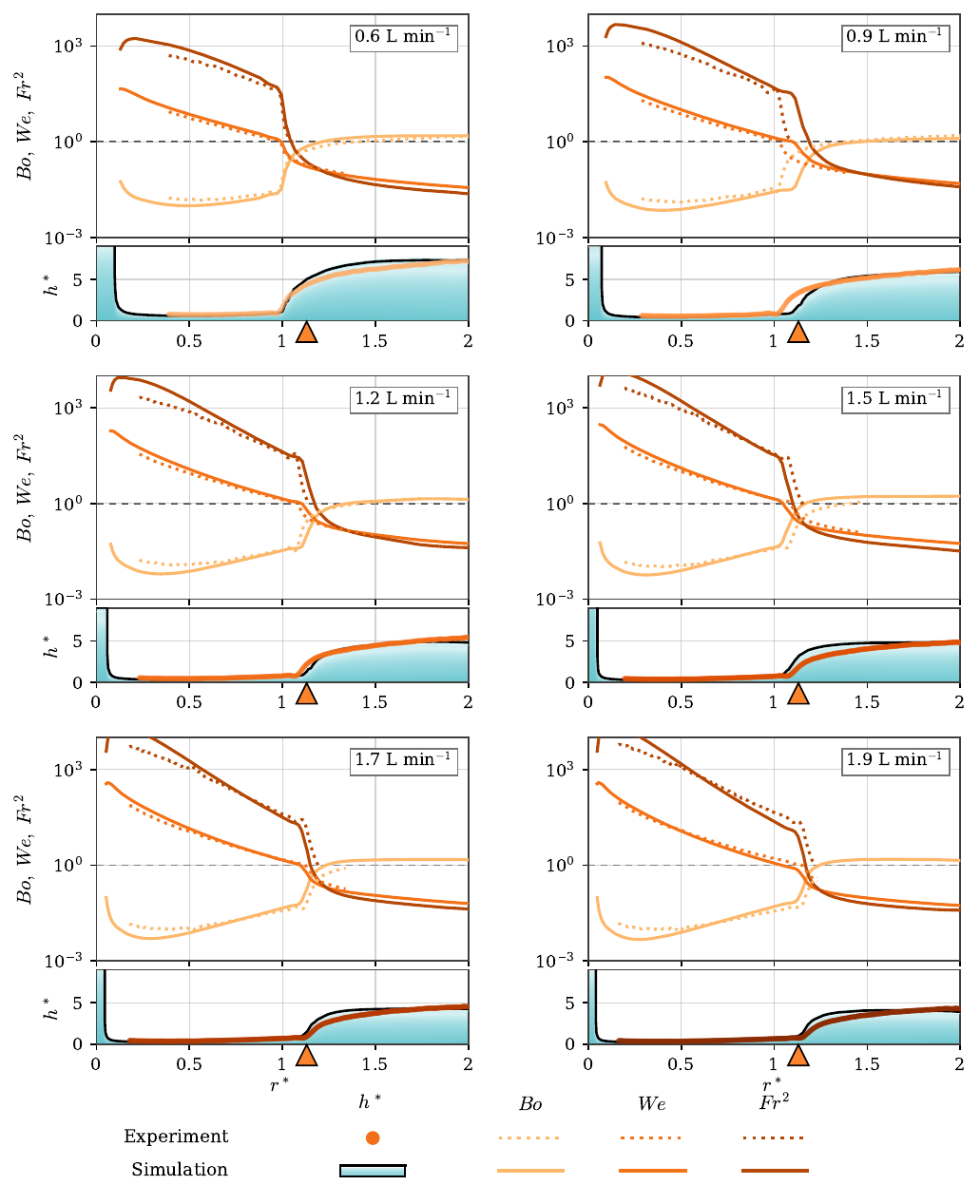}

 \caption{Comparison of the experimental (markers) and numerical (black line) film profiles for the \pfl{water-glycerine mixture} WG~8/7 at flow rates $Q=\SIrange{0.6}{1.9}{\litre\per\minute}$ under steady conditions. The radius is scaled  using $R_\gamma$~\eqref{eq:Rst} and film thickness using corresponding capillary scale \(\delta_{\gamma}=\left(\rho\nu q/\gamma\right)^{1/2}\) (appendix~\ref{eq:capillary_scales}). The lower panels compare the measured and simulated height profiles, while the upper panels show the corresponding $Bo$, $We$ and $Fr^2$ obtained from the simulations (solid lines) and experiments (dotted lines). Overall, the simulations agree well with the experiments. At $Q=\SI{1.7}{\litre\per\minute}$ and $\SI{1.9}{\litre\per\minute}$, the simulated jump is sharper, with a post-jump height discrepancy of approximately $\SIrange{0.5}{0.7}{\milli\metre}$, possibly due to optical lensing in the curved jump region. In the thin-film region, $Bo=O(10^{-2})$--$O(10^{-1})$, while near the jump $We=O(1)$ and $Fr^2\gg1$, consistent with a capillary-dominated jump}
    \label{fig:film_thickness sim vs expt}
\end{figure}

The depth profiles obtained from the numerical simulations are compared with the corresponding experimental measurements in figure~\ref{fig:film_thickness sim vs expt} for the WG 8/7 mixture over the flow-rate range $Q=\SIrange{0.6}{1.9}{\litre\per\minute}$. The corresponding dimensionless flow rate $Q^*$ ranges from 11.8 to 37.4. \rkb{The radial coordinate is made dimensionless with the capillary scale
\(R_{\gamma}\) defined in \eqref{eq:Rst}, and the film thickness with
the corresponding capillary thickness scale,
\(\delta_{\gamma}=\left(\rho\nu q/\gamma\right)^{1/2}\)} (appendix~\ref{eq:capillary_scales}). The lower panels show the measured and simulated film profiles, while the upper panels present the corresponding $Bo$, $We$ and $Fr^2$ calculated directly from the numerical results and compared with those obtained from the experiments. The local $We$ and $Fr^2$ are calculated using the depth-averaged velocity. The results are also compared with calculations based on the film profile weighted velocity (the momentum flux weighted \(\bar{u^2} = 1/h \int u^2 dz \)), and no significant difference is observed between the two formulations. 

\rkb{The numerical and experimental film profiles are in good agreement. The simulations reproduce both the supercritical film thickness upstream of the jump and the jump radius across the full range of flow rates. The predicted post-jump film thickness also follows the experimental trend, demonstrating that the numerical model
captures the principal features of both the supercritical and subcritical regions. The corresponding triangle marks the prediction
\(R_j/R_\gamma=1.13\) from \eqref{eq:rc}. Since this far-field scaling neglects the boundary-layer formation region, a slight mismatch is
expected at the lowest flow rate in panel (\textit{a}). Nevertheless,
\eqref{eq:rc} provides excellent agreement with the experimental jump radii.}

The orders of magnitude of the governing parameters are consistent with those obtained experimentally and shown in figure \ref{fig:water_WG_W2P_Bo_Fr_We}. This behaviour further supports the previous conclusion that the jump is controlled by capillary effects rather than gravity. 

\subsection{Effect of the surface tension}
\label{sec:surface_tension}

The strictest numerical test of the proposed mechanism is to vary the surface tension alone while leaving the density, viscosity and flow rate unchanged, because the capillary and gravitational descriptions differ explicitly in their dependence on \(\gamma\). The capillary scaling \eqref{eq:Rst} predicts \(R_j\propto\left(\frac{\rho Q^3}{\nu\gamma}\right)^{1/4},\) so that reducing the surface tension should move the jump outwards by a prescribed amount. The gravity-wave criterion \eqref{eq:Rg}, by contrast, predicts a critical radius independent of surface tension.

\begin{figure}
    \centering
    \includegraphics[width=1\linewidth]{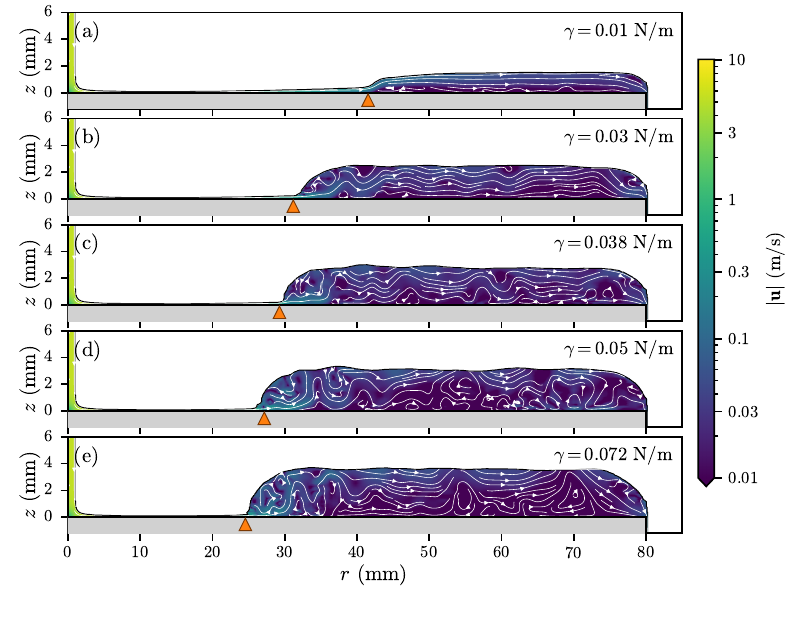}
    \caption{Steady-state film profiles produced by the impingement of a liquid jet ($\rho = \SI{1000}{\kilogram\per\meter\cubed}$,$\nu = \SI{1e-6}{\meter\squared\per\second}$)  at a fixed flow rate $Q = \SI{1}{\litre\per\minute}$. The surface tension alone is varied: (a) $\SI{0.01}{\newton\per\meter}$ (b) $\SI{0.03}{\newton\per\meter}$, (c) $\SI{0.038}{\newton\per\meter}$ (Water+SDBS solution \citealp{bhagat2018origin}), (d) $\gamma = \SI{0.05}{\newton\per\meter}$   (e) $\gamma = \SI{0.072}{\newton\per\meter}$ (water). The triangles represent the locations of the predicted jump radius from surface tension-based scaling in equation \eqref{eq:rj}.} Increasing the surface tension decreases the jump radius from $R_j = \SI{41}{\milli\meter}$ in (a) to
$\SI{24}{\milli\meter}$ in (e).
\label{fig:surface_tension}
\end{figure}

The computations were performed at fixed
\(\rho=1000\,\mathrm{kg\,m^{-3}}\), \(\nu=1\times10^{-6}\,\mathrm{m^2\,s^{-1}}\) and
\(Q=1\,\mathrm{L\,min^{-1}}\), with \(\gamma=0.010\), \(0.030\), \(0.038\), \(0.050\) and \(0.072\,\mathrm{N\,m^{-1}}\). The value \(\gamma=0.038\,\mathrm{N\,m^{-1}}\) corresponds to the water--SDBS experiments of \citet{bhagat2018origin}, while \(\gamma=0.072\,\mathrm{N\,m^{-1}}\) corresponds to pure water. Figure~\ref{fig:surface_tension} shows the resulting steady flow fields.
At the lowest surface tension the downstream film is comparatively thin and contains little recirculation. As \(\gamma\) is increased, the jump moves progressively inwards, the subcritical film deepens, and increasingly strong recirculating structures develop downstream. The jump location, however, remains sharply defined in every case. (See Supplementary video 3)

\begin{figure}
    \centering
    \includegraphics[width=1\linewidth]{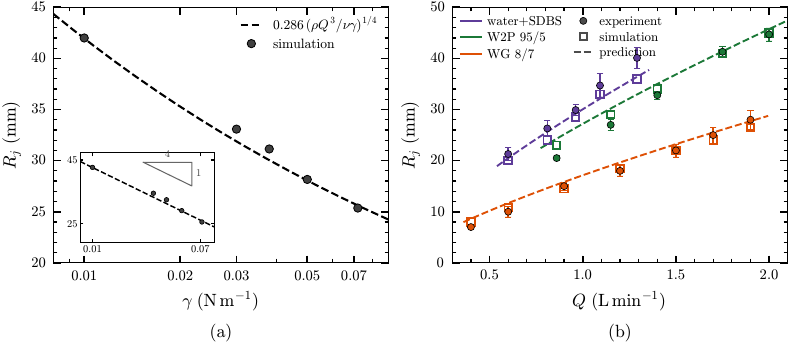}
    \caption{The effect of surface tension on the jump radius. (\textit{a})~Simulated jump radius as a function of surface tension
for the model fluids, which share the density and viscosity of water, at \(Q=1\,\mathrm{L\,min^{-1}}\): only \(\gamma\) is varied. The dashed line is the prediction \eqref{eq:rj}. The
inset shows the same data on logarithmic axes, on which the triangle indicates the slope \(-1/4\). (\textit{b})~Measured jump radii \citep{bhagat2018origin} (filled symbols, error bars \(\pm\) one standard deviation) and simulated (open symbols) jump radii as functions of flow rate for water+SDBS, W2P~95/5 and WG~8/7. The dashed lines are \eqref{eq:rj} evaluated using the appropriate physical properties of each case.}
\label{fig:surface_tension2}
\label{fig:Rj vs gamma}
\end{figure}

The jump radius decreases monotonically from
\(42.7\,\mathrm{mm}\) at
\(\gamma=0.010\,\mathrm{N\,m^{-1}}\) to
\(25.3\,\mathrm{mm}\) at
\(\gamma=0.072\,\mathrm{N\,m^{-1}}\). Thus a factor of \(7.2\)
increase in surface tension reduces the jump radius by a factor of
\(1.69\), compared with the factor \(7.2^{1/4}=1.64\) required by the
\(\gamma^{-1/4}\) dependence. 
The triangles in figure~\ref{fig:surface_tension} mark the corresponding predicted positions given by \eqref{eq:rj}. \rkb{Figure~\ref{fig:surface_tension2}(\textit{a}) shows the simulated jump
radius as a function of surface tension. The simulated radii lie within \(4\,\%\)
of the parameter-free prediction \eqref{eq:rj}, drawn with no fitted
quantity. In the inset the same data are shown on logarithmic axes, on which the prediction is a straight line of slope \(-1/4\).} Surface tension alone is changed, yet the jump moves by a factor of approximately
\(1.7\), in both the direction and the magnitude predicted by the capillary scaling. 

\rkb{Figure~\ref{fig:surface_tension2}(\textit{b}) extends this comparison to the flow-rate dependence for the experimental liquids, comparing experiment, simulation and scaling directly in dimensional variables. For each liquid, the measured jump radii (filled symbols) and the simulated ones (open symbols) are plotted against flow rate, together with the prediction \eqref{eq:rj} evaluated using the physical properties of each liquid (dashed lines). \citet{bhagat2018origin} measured the jump radius for water and for a saturated water--SDBS solution, whose density and viscosity
are essentially those of water while \(\gamma\) is nearly halved, from
\(0.072\) to \(0.038\, \SI{}{\newton\per\meter}\). The jump radius increased, and \eqref{eq:rj} places the increase at \((0.072/0.038)^{1/4}=1.17\), i.e.\ \(17\,\%\), at the same flow rate.
An increase of the jump radius upon the addition of surfactant was also noted by \citet{bush2003influence}. The second fluid of the same study, W2P~95/5, has a slightly higher viscosity than water
(\(1.27\times10^{-6}\) against
\(1.02\times10^{-6}\, \SI{}{\meter\squared\per\second}\)) and a lower surface
tension (\(0.0425\,\SI{}{\newton\per\meter}\)): increasing the viscosity
reduces the jump radius while reducing the surface tension increases
it, and \eqref{eq:rj} predicts the net of these two opposing
quarter-power shifts, a jump \(8\,\%\) beyond water. WG~8/7, with
nearly six times the viscosity of water, is predicted at \(0.67\) of the 
water radius despite its comparable surface tension. The simulations
reproduce each of these changes, and measured and simulated
radii alike follow their corresponding parameter-free predictions
across the flow-rate range. Thus the controlled simulations isolate
the effect of surface tension, while the experimental comparisons show
that \eqref{eq:rj} also captures the combined dependence on the fluid
properties.}

\subsection{Effect of the contact angle}
\label{sec:contact_angle}

\begin{figure}
    \centering
\includegraphics[width=\linewidth]{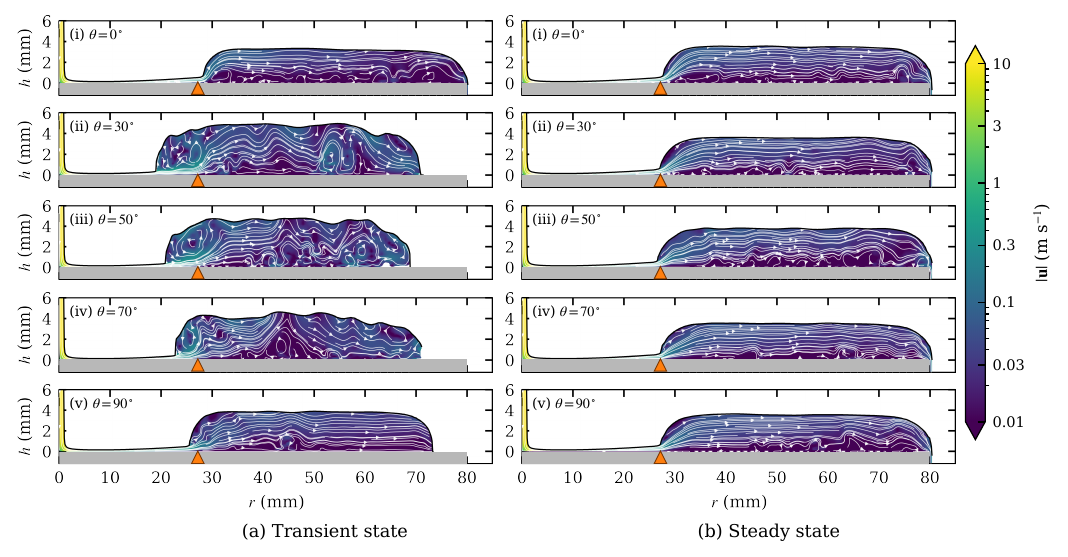}

\caption{Effect of the imposed plate contact angle on the circular hydraulic jump for WG~8/7 at \(Q=1.9\,\mathrm{L\,min^{-1}}\), shown
(a) during the transient spreading at \(t=1.7\,\mathrm{s}\) and (b) at steady state. Each column contains five cases,
\(\theta=0^\circ\), \(30^\circ\), \(50^\circ\), \(70^\circ\) and \(90^\circ\). The triangles mark the contact-angle-independent jump radius predicted by \eqref{eq:rj}, \(R_j=27.2\,\mathrm{mm}\). Although the imposed
contact angle strongly alters the transient spreading and downstream flow structure, all five cases converge to the same steady jump location.}
    \label{fig: effect of CA}
\end{figure}

To test directly whether the wetting condition of the advancing liquid influences the jump location, the WG~8/7 computation at \(Q=1.9\,\mathrm{L\,min^{-1}}\) was repeated for imposed plate contact
angles \(\theta=0^\circ\), \(30^\circ\), \(50^\circ\), \(70^\circ\) and \(90^\circ\), with all other parameters unchanged. Despite the
large variation in the  flow after the jump, all five cases converge to  the same steady jump radius, \(R_j\simeq27.5\pm0.5\,\mathrm{mm}\), in agreement with the
contact-angle-independent prediction of \eqref{eq:rj}. The triangles in
figure~\ref{fig: effect of CA} mark this predicted position. The
significance of the result is apparent from the transient calculations:
changing the contact angle substantially alters the spreading and
structure of the subcritical film while leaving the jump itself fixed.

Figure~\ref{fig: effect of CA}(a) shows the five cases at
\(t=1.7\,\mathrm{s}\), while the outer film is still spreading. Both
the extent of the film and the character of the downstream flow depend
strongly, but non-monotonically, on \(\theta\). For
\(\theta=0^\circ\), the perfectly wetting film spreads most rapidly and
has already reached the plate rim by this time, so that no advancing
contact line remains on the plate. The other four films have advanced
only to approximately \(68\)--\(73\,\mathrm{mm}\). At
\(\theta=90^\circ\), the advancing front instead forms a pronounced
bead and progresses more slowly but comparatively steadily. At
intermediate contact angles the free surface of the subcritical film
develops substantial vertical motion and a recirculating roller forms
above the jump, while the high-speed stream continues beneath it as an
underflow. Such rollers are a documented feature of circular hydraulic
jumps \citep{craik1981circular,liu1993hydraulic,bush2006experimental}.
The roller can overhang the thin film upstream, so that the apparent
leading edge of the thick region lies inside the jump radius. These
marked differences in the transient downstream state disappear and every case converges to the same steady jump
position (see Supplementary video 4 ).

The steady-state result in figure~\ref{fig: effect of CA}(b) is
consistent with the experiments of
\S\ref{Sec: experimental observation}, in which the jump radius remains
unchanged while the contact line of the spreading film advances and
ultimately disappears. It is also consistent with circular jumps
observed on plates at different orientations
\citep{bhagat2018origin} and with film jumps on surfaces having
different wetting characteristics \citep{bhagat2016flow}.

The intermediate contact angles also produce the strongest vertical
and recirculating motion in the flow outside the jump. As a simple measure
of this motion, we consider
\(\overline{|u_z|/|\mathbf{u}|}\), evaluated in the subcritical liquid
region. It is \(0.16\) for \(\theta=0^\circ\), rises to \(0.33\) at
\(\theta=50^\circ\), and falls to \(0.12\) at \(\theta=90^\circ\).
The non-monotonic variation suggests that the disturbance is associated
with the wetting condition rather than simply with the low velocity of
the downstream film. The observed behaviour is consistent with
intermittent motion of the advancing front, in which liquid accumulates
behind a slowly advancing contact line and is subsequently released,
driving the recirculating structures.

The origin of this transient behaviour cannot, however, be established
conclusively from the present calculations. The simulations impose a
static contact angle, whereas an advancing contact line generally
exhibits a dynamic contact angle and may also display contact-angle
hysteresis \citep{wang2015flow}. Moreover, the slowly moving,
strongly curved subcritical film is the part of the calculation most
susceptible to parasitic numerical currents
(\S\ref{sec:num_protocol}). These qualifications concern the transient
outer film rather than the principal result: the steady jump location
is insensitive to the imposed contact angle.

\subsection{Jump-radius scaling across the full data set}
\label{sec:jump_radius_collapse}
 
\begin{figure}
    \centering
    \includegraphics[width=\linewidth]{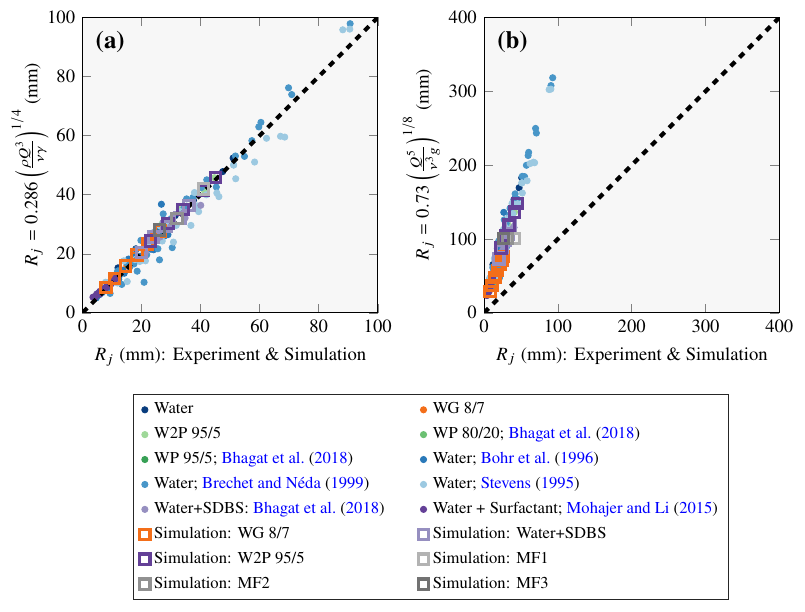}
    \caption{Comparison of the measured and simulated jump radii with theoretical predictions. The data span flow rates $Q = 0.097$--$\SI{6.3}{\liter\per\minute}$, surface tensions $\gamma = 0.01$--$\SI{0.072}{\newton\per\meter}$, and kinematic viscosities $\nu = 1\times10^{-6}$-- $6\times10^{-6} \SI{}{\meter^2\per\second}$. The figure contains 203 experimental data points and 19 numerical data points, giving 222 points in total, with $Q^* = 3.84$--$1775.56$. Panel (a) compares the data with the surface-tension-based prediction from equation \eqref{eq:rj}, while panel (b) compares them with the gravity-based scaling of \cite{bohr1993shallow}. The measurements and simulations agree well with the surface-tension-based prediction, whereas the gravity-based scaling substantially overpredicts the jump radius.}
    \label{fig: jump radius scaling vs expt}
\end{figure}

 \begin{table}
\caption{Fluid properties and flow-rate ranges from present and previous studies.}
\label{tab:fluid properties of present and previous studies}
\centering
\begin{tabular}{ccccccc}
\hline
\textbf{Fluid} & \textbf{$\rho$ ($\SI{}{\kilogram \meter^{-3}}$)} & \textbf{$\nu$ ($\SI{}{\meter^2 \second^{-1}}$)} & \textbf{$\gamma$ ($\SI{}{\newton \meter^{-1}}$)}& Q $(\SI{}{\liter \per \minute})$ & \textbf{$Q^*$} & \textbf{Reference} \\
\hline
Water     & $1000$ & $1.02 \times  10^{-6}$ & 0.072 & 0.11 - 6.31 & 30.95--1775.56 & \makecell{\cite{bohr1993shallow};\\ \cite{brechet1999circular}; \\ \cite{stevens1993measurements}}\\ 

\makecell{Water Propanol\\ (WP 95/5)} & $989$ & $1.27 \times 10^{-6}$ & 0.0425 & 0.5 - 1 & 55.88 - 111.76 & \cite{bhagat2018origin} \\ 
\makecell{Water Propanol\\ (WP 80/20)} & $968$ & $2.3 \times 10^{-6}$ & 0.026 & 0.38 - 1.15 & 10.48 - 31.72 & \cite{bhagat2018origin} \\ 

Water+Surfactant & $1000$ & $1 \times 10^{-6}$ & 0.037 & 0.097 - 0.3 & 175.37 - 542.37 &  \cite{mohajer2015circular} \\ \\
Water+SDBS & $1000$ & $1 \times 10^{-6}$ & 0.038 & 0.6 - 1.3 & 42.69 - 92.47 &  \cite{bhagat2018origin} \\ \\

\makecell{Water Glycerine \\ (WG 8/7)}      & $1132$ & $6 \times  10^{-6}$ & 0.067 & 0.6 - 1.9 & 11.81--37.4 & This study\\ 

MF1 & $1000$ & $1 \times 10^{-6}$ & 0.01 & 1 & 3.84 & This study \\ \\
MF2 & $1000$ & $1 \times 10^{-6}$ & 0.03 & 1 & 34.58 & This study \\ \\
MF3 & $1000$ & $1 \times 10^{-6}$ & 0.05 & 1 & 96 &  This study \\ 

\end{tabular}
\end{table}
Having established the critical condition from the film profiles, we finally test the resulting jump-radius scaling against the full
experimental and numerical data set, \pfl{including all published data that satisfy $Q^* > 1$}. Figure~\ref{fig: jump radius scaling vs expt} compares 203 experimental measurements and 19 simulated jump radii, giving 222 cases in total. The data span the fluids listed in table~\ref{tab:fluid properties of present and previous studies}, with
\(\nu=\SIrange{1e-6}{6e-6}{\metre\squared\per\second}\),
\(\gamma=\SIrange{0.01}{0.072}{\newton\per\metre}\),
\(\rho=\SIrange{968}{1132}{\kilogram\per\metre\cubed}\),
jet diameters from \(\SI{2.2}{\milli\metre}\) to
\(\SI{8.9}{\milli\metre}\), and \(Q=\SIrange{0.097}{6.3}{\litre\per\minute}\). The corresponding
dimensionless flow rate spans \(Q^*=3.84\)--\(1775\).

Figure~\ref{fig: jump radius scaling vs expt}(\textit{a}) compares the measured and simulated jump radii with the leading-order capillary prediction \eqref{eq:rj}, while
figure~\ref{fig: jump radius scaling vs expt}(\textit{b}) compares the same data with the gravity-based scaling. In both panels the black dashed line denotes equality between the observed and predicted jump radii. Under the capillary scaling the data cluster closely about this line over the full range of fluid properties and flow conditions. Under the gravity scaling, by contrast, the predicted radii lie predominantly above the line of equality, showing that the gravitational scaling systematically overpredicts the observed jump radius.

The comparison separates the two questions. Watson's similarity solution
describes the radial development of the upstream film under either
normalisation, so the film is viscous; but only the capillary scale
\(R_{\gamma}\) collapses the position of the jump, so the jump is selected
when the local capillary-critical condition \(We=O(1)\) is reached,
consistent with the direct measurement of
\S\ref{Sec: experimental observation}. The inertia--viscous balance
underpinning both scalings is thus established independently by the
measurements and by the governing equations; of the two closures,
only the capillary one is consistent with the data. In this regime, surface tension, not gravity, sets the circular hydraulic jump.
\section{Conclusion}
\label{sec:conclusion}
\rkb{
This paper has examined the circular hydraulic jump at the scale of a kitchen sink using experiments, scaling analysis and numerical
simulations. Experiments with water, water--2-propanol and water--glycerine solutions provided direct measurements of the film thickness throughout the supercritical and subcritical regions. During the transient \pfl{development of the flow}, the jump becomes stationary while the outer film continues to spread and deepen, reaches the plate rim and ultimately drains continuously into the surrounding pool, leaving no contact line
on the plate. Over this evolution the outer boundary moves by nearly \(70\,\mathrm{mm}\), while the jump remains fixed. Thus substantial
changes in both the downstream state and its boundary condition occur without displacing the jump, demonstrating that the jump location is
established independently of the evolving outer film.
 
The measured film thickness, together with volume-flux conservation, gives the local depth-averaged radial velocity and hence the local
\(Bo\), \(We\) and \(Fr^2\). Throughout the supercritical film, \(Bo=O(10^{-3})\)--\(O(10^{-1})\). At the measured jump radii,
\(Fr^2=O(10)\)--\(O(10^2)\) and varies substantially between fluids and flow rates, and there is no common Froude number associated with
the jump. In contrast, the local Weber number is \(We=O(1)\) in every case. The jump therefore occurs while the film remains strongly supercritical with respect to gravity waves, but at a common capillary condition.
 
The dimensional analysis identifies the competing capillary and gravitational radial scales and the dimensionless flow rate \(Q^*\)
that distinguishes the corresponding regimes. In the present regime the developed upstream film is inertia--viscous. Closing this film
with the capillary-critical condition \(We=1\) gives \(R_j=0.286(\rho Q^3/\nu\gamma)^{1/4}\), equation~\eqref{eq:rj}. The prefactor is only weakly sensitive to the
assumed velocity shape function: the cubic shape function gives \(0.281\), only \(1.7\,\%\) below the Watson value, while the parabolic shape function gives \(0.264\), \(7.7\,\%\) below it; the scaling with the flow parameters is unchanged.
 
The numerical calculations carried out under matched experimental conditions reproduce the measured supercritical film thickness and
jump radius and recover the same local ordering near the jump: \(Bo\ll1\), \(We=O(1)\) and \(Fr^2\gg1\).
 
The simulations then provide two controlled tests in which individual parameters are varied independently. First, keeping \(\rho\), \(\nu\)
and \(Q\) fixed while increasing \(\gamma\) from
\(0.010\) to \(0.072\,\mathrm{N\,m^{-1}}\) decreases the jump radius from \(42.7\) to \(25.3\,\mathrm{mm}\), a factor of \(1.69\), compared with the factor \(7.2^{1/4}=1.64\) predicted by \(R_j\propto\gamma^{-1/4}\). Surface tension alone therefore moves the
jump in both the direction and magnitude predicted by the capillary scaling.
 
Second, varying the imposed contact angle from \(0^\circ\) to \(90^\circ\), with all other parameters unchanged, substantially changes the spreading, thickness and free-surface structure of the subcritical film and produces markedly different downstream flow states, including a recirculating roller at intermediate contact
angles. Nevertheless, all five cases converge to essentially the same steady jump radius, \(R_j\simeq27.5\pm0.5\,\mathrm{mm}\). Together with the transient experiments, in which the contact line advances across the plate and ultimately disappears without moving the
jump, this establishes that the wetting condition does not select the jump radius.
 
The resulting capillary scaling was finally tested against the full data set of 203 experimental and 19 numerical jump radii. These experimental data set includes the experiments conducted in this study and the experimental data available in literature. These 222 cases span \(Q=0.097\)--\(6.3\,\mathrm{L\,min^{-1}}\),
\(\gamma=0.01\)--\(0.072\,\mathrm{N\,m^{-1}}\), a sixfold range of kinematic viscosity and \(Q^*=3.84\)--\(1775\). The measured and
simulated radii cluster closely about the parameter-free capillary prediction over this entire range, whereas the gravity-based scaling
systematically overpredicts the observed jump radius.
 
These observations carry a broader conceptual conclusion: the hydraulic jump is not merely a force balance. Once formed, the jump must satisfy a local momentum balance, but that balance does not by itself determine where the transition occurs. Hydraulic control is established when the supercritical film reaches its critical condition, arresting upstream communication; the downstream film then adjusts to the resulting transition. Consequently, a change in downstream film depth, wetting condition or contact-line position cannot, by itself, establish that a downstream force balance selects the jump. The discriminating test is whether the jump position follows the
capillary-critical prediction as \(\gamma\) is varied while remaining unexplained by the gravity-based one --- and this is what is observed.
 
Gravity is not absent from the flow: it influences the depth and drainage of the subcritical outer film. The scaling analysis
identifies \(Q^*=Q_c/Q\), with \(Q^*\sim1\) marking the crossover between the capillary- and gravity-influenced regimes. The present
experiments lie entirely in the \(Q^*>1\) regime, where the jump is reached while \(Fr^2\gg1\) and \(We=O(1)\). The present conclusion therefore does not imply that gravity can never determine a circular hydraulic jump, but identifies the regime in which it does not.
 
For the circular hydraulic jump at the scale of a kitchen sink, the combined experimental and numerical evidence admits one conclusion:
surface tension, not gravity, sets the jump.}

\section*{Acknowledgements}
RKB would like to acknowledge funding from The Royal Academy of Engineering (RF2122-21-305).   
\section*{Declaration of Interests}
The authors report no conflict of interest.
\appendix
\section{Dye Calibration}
\label{append: dye calibration}

\begin{figure}
      \begin{subfigure}[h]{0.59\linewidth}
      \centering
        \includegraphics[width=1\linewidth]{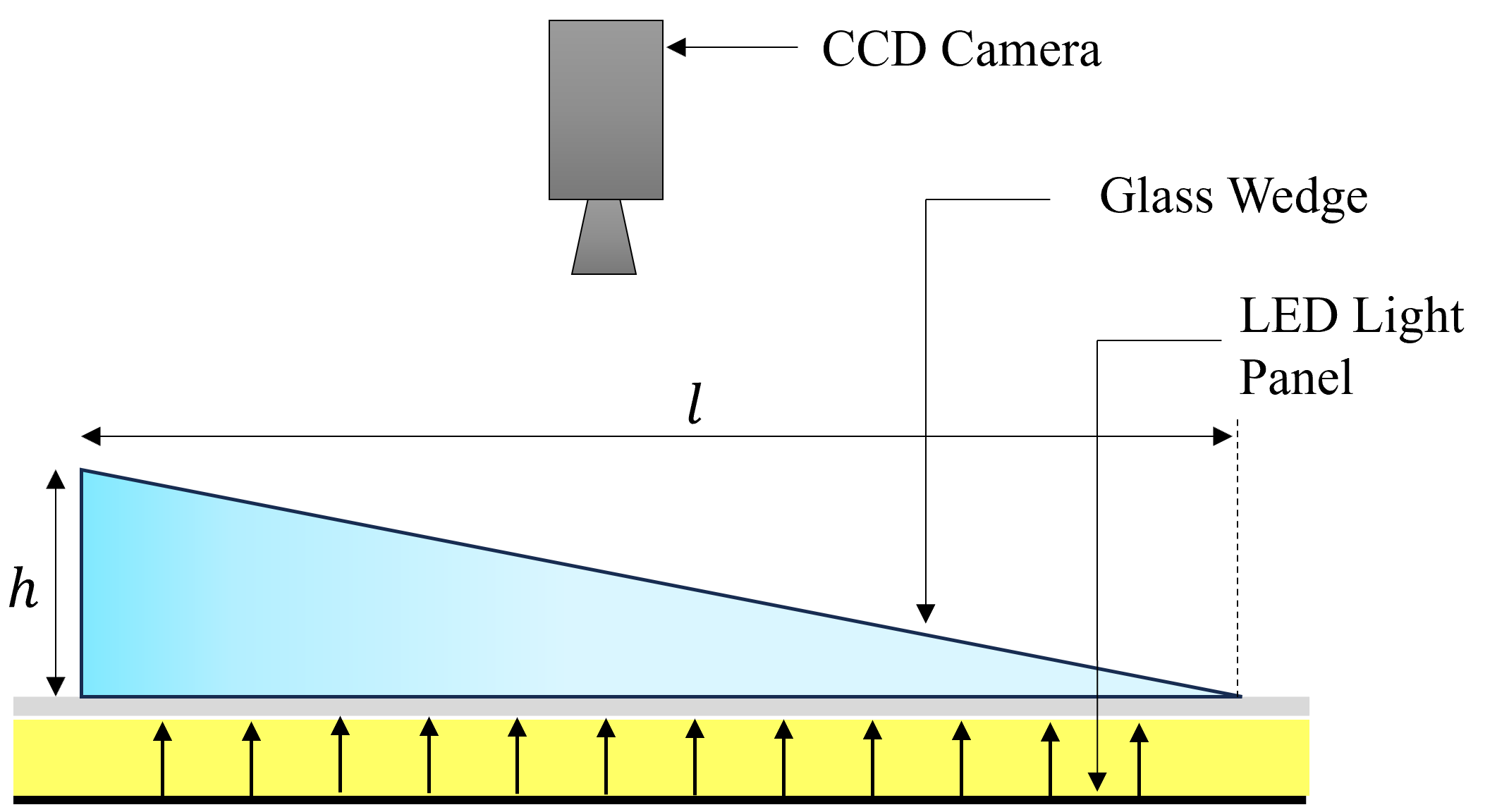}
      \caption{}
      \end{subfigure}
   \begin{subfigure}[h]{0.4\linewidth}
      \centering
        \includegraphics[]{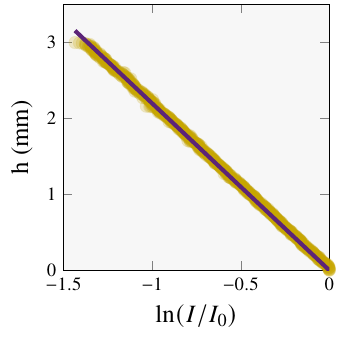}
      \caption{}
      \end{subfigure}
  \caption{(a) shows the front view of the dye calibration setup. The wedge filled with water nigrosin dye is evenly illuminated from below by a LED light panel. The image of the wedge is taken from above using a CCD camera. Figure (b) shows the calibration curve for film thickness estimation given by equation \ref{Beer-Lambert law} where $\epsilon = - 2.198$ } 
  \label{Fig: dye cal}
\end{figure}

 The absorption of light passing through the dye (nigrosin) solution is governed by the Beer–Lambert law, which depends on the dye concentration and the optical path length, or liquid film thickness \citep{kim2005photochromic}. This relationship can be mathematically expressed as,
\begin{equation}
\label{Beer-Lambert law}
h = \epsilon log(I/I_0),
\end{equation}
where, $h$ is the thickness of the film, $\epsilon$ is the fluid absorptivity, $I_0$ is the background light intensity, and $I$ the light intensity after absorption or passing through the film. Dye calibration was achieved by capturing images of liquid films of known thickness and the background, resulting in $I_o$ and $I$. To generate liquid films of known thickness, \cite{wang2017drop} enclosed a known volume of liquid between two glass slides separated by coverslips, thereby creating a liquid film of known thickness. In this study, we devised a straightforward calibration procedure that traps liquid within a wedge made of glass slides, as illustrated in figure \ref{Fig: dye cal}(a). The liquid trapped inside the hollow wedge enable us to establish a light path length that decreases linearly, ranging from the height of the wedge to zero, thus producing a continuous calibration curve with a resolution of 0.005 \SI{}{\milli\meter}. The image of the empty wedge is used to determine the background light intensity $I_o$, while the wedge filled with liquid provides the intensity $I =I(h)$, where $h$ is the height of the wedge or, the height of the trapped liquid film. The intensity ratio, $I/I_0$, is calculated by dividing the images pixel by pixel in MATLAB. Figure \ref{Fig: dye cal}(b) shows a plot for the height vs the intensity ratio for the wedge. Before each run, the calibration of nigrosin-water is carried out to obtain the relation between light intensity ratio ($I/I_0$) and film thickness ($h$). 


\section{Experimental uncertainties and data smoothing}
\label{append: data smoothing}
We accounted for the various sources through which errors may propagate and calculated the combined uncertainty in the measured film height. One source of uncertainty arises from measuring the dye mass using a weighing scale with a resolution of $\pm\SI{0.001}{\gram}$. Additional uncertainties arise from measuring the height and length of the calibration wedge, both of which were measured using a scale with a resolution of $\pm\SI{0.1}{\milli\metre}$. The wedge height varies linearly along its length according to $h=\frac{h_w}{l_w}x$, where $h_w$ is wedge height and $l_w$ is the wedge length. Assuming that the uncertainties in dye mass, $h_w$ and $l_w$ are independent, the relative uncertainty in the measured film height can be estimated as \citep{moffat1988describing},
\begin{equation}
\frac{\Delta h}{h} = \sqrt{\left(\frac{\Delta m}{m}\right)^2 + \left(\frac{\Delta h_w}{h_w}\right)^2 + \left(\frac{\Delta l_w}{l_w}\right)^2}. \label{eq:height_uncertainty} 
\end{equation} 
\begin{figure}
    \centering
    \includegraphics[width =0.7\linewidth]{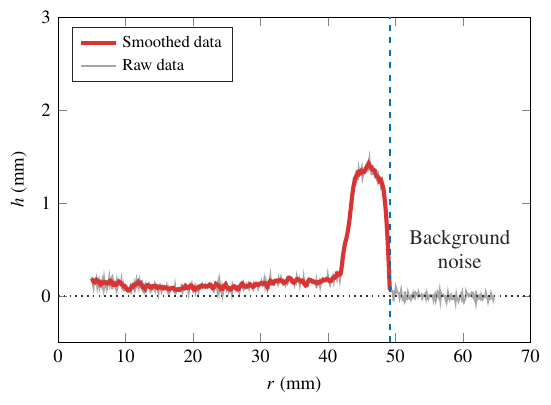}
    \caption{Comparison of the raw and smoothed film-thickness profiles during jump formation at $t=\SI{0.12}{\second}$. The data beyond the blue dashed line correspond to the background region and fluctuate about zero. Based on the magnitude and length scale of these fluctuations, a Savitzky–Golay filter with a nine-point window and a second-order polynomial was used to smooth the measured profile while preserving the sharp variation across the jump.}
    \label{fig:raw vs smooth data Savitzky–Golay filter}
\end{figure}

The measured film-thickness profiles were smoothed using a Savitzky–Golay filter \citep{savitzky1964smoothing,schafer2011savitzky}. The filter replaces each point with the value obtained from a local least-squares polynomial fit to the neighbouring data. A window length of nine points was selected by examining the fluctuations in the background measurements as shown in figure \ref{fig:raw vs smooth data Savitzky–Golay filter}. This window is sufficiently large to suppress noise associated with the camera sensor and illumination while preserving the sharp change in film thickness across the hydraulic jump. A second-order polynomial was used because the film-thickness profile is locally curved rather than purely linear. The quadratic fit therefore preserves the local slope and curvature of the profile more accurately than a moving average or linear fit, without introducing the additional sensitivity to noise associated with higher-order polynomials.
\section{Scaling of the viscous shallow-water equation}
\label{app:film_scaling}

We work in cylindrical polar coordinates \((r,\theta,z)\), with \(u\)
and \(w\) the radial and wall-normal velocity components, and introduce
a radial length scale \(R\), a film-thickness scale \(\delta\), and a
radial-velocity scale \(\upsilon\), writing the dimensionless variables
with an asterisk,
\begin{equation}
    r=Rr^{*},
    \qquad
    z=\delta z^{*},
    \qquad
    h=\delta h^{*},
    \qquad
    u=\upsilon u^{*},
    \qquad
    w=\epsilon\upsilon w^{*},
    \qquad
    \epsilon=\frac{\delta}{R}\ll1.
\label{eq:film_scaling_variables}
\end{equation}
Volume conservation relates the scales through
\begin{equation}
    q=Urh\sim\upsilon R\delta,
    \qquad q=\frac{Q}{2\pi},
\label{eq:volume_scaling}
\end{equation}
and the viscous shallow-water (radial boundary-layer) equation
\citep{tani1949water}, away from the narrow jump region, becomes
\begin{equation}
    u^{*}\frac{\partial u^{*}}{\partial r^{*}}
    +w^{*}\frac{\partial u^{*}}{\partial z^{*}}
    =-\frac{1}{Fr^{2}}\frac{\mathrm{d}h^{*}}{\mathrm{d}r^{*}}
    +\frac{1}{\epsilon Re}
    \frac{\partial^{2}u^{*}}{\partial z^{*2}}
    +O(\epsilon^{2}),
    \qquad
    Re=\frac{\upsilon\delta}{\nu},
    \quad
    Fr=\frac{\upsilon}{\sqrt{g\delta}},
    \quad
    We=\frac{\rho\upsilon^{2}\delta}{\gamma}.
\label{eq:nondim_radial_momentum}
\end{equation}
The \(O(\epsilon^{2})\) remainder comprises the radial viscous
diffusion and the streamwise capillary pressure gradient, the latter
being of order \(\epsilon^{2}/We\). Both pressure gradients are
subdominant: the hydrostatic gradient because the measurements show
the developed upstream film to remain strongly gravity-supercritical,
\(Fr^{2}\gg1\) (\S\ref{Sec: experimental observation}) --- its neglect
therefore rests on the measured regime, not on an assumed gravity-free
limit --- and the capillary gradient because \(\epsilon\ll1\), even
where \(We\) falls to \(O(1)\).

Retaining both inertia and viscous diffusion at leading order requires
\(\epsilon Re=O(1)\), i.e.\ the inertia--viscous balance
\(\upsilon\delta^{2}\sim\nu R\); the leading-order equation is then
\eqref{eq:parameter_free} of \S\ref{subsec:jump_location}, which in
the scaled variables contains no parameter. This balance and the flux
relation \eqref{eq:volume_scaling} fix two of the three scales; the
closing relation \(We=O(1)\), i.e.\
\(\rho\upsilon^{2}\delta\sim\gamma\), fixes the third:
\begin{equation}
    R_{\gamma}=\left(\frac{\rho q^{3}}{\gamma\nu}\right)^{1/4},
    \quad
    \delta_{\gamma}=\left(\frac{\rho\nu q}{\gamma}\right)^{1/2},
    \quad
    \upsilon_{\gamma}
    =\left(\frac{\gamma^{3}}{\rho^{3}q\nu}\right)^{1/4},
    \quad
    \epsilon_{\gamma}
    =\left(\frac{\rho\nu^{3}}{\gamma q}\right)^{1/4},
    \quad
    Re_{\gamma}
    =\frac{\upsilon_{\gamma}\delta_{\gamma}}{\nu}
    =\frac{1}{\epsilon_{\gamma}},
\label{eq:capillary_scales}
\end{equation}
so that
\begin{equation}
    \epsilon Re=1
\label{eq:epsilon_Re_unity}
\end{equation}
identically: the capillary scales satisfy the inertia--viscous balance
exactly. For the experimental parameters the resulting aspect ratios
are \(\epsilon\approx0.007\)--\(0.03\), consistent with the thin-film
ordering.

 \section{Mesh}
 \label{sec:mesh}
 
The computational domain was divided into two blocks in the $y$-direction: a near-wall region extending from $y=0$ to $\SI{1}{\milli\metre}$ and an outer region extending from $y=\SI{1}{\milli\metre}$ to $\SI{10}{\milli\metre}$. A uniform grid spacing of $\SI{0.04}{\milli\metre}$ was used in the near-wall block to resolve the strong velocity gradients and the thin liquid film ($0.2 - \SI{0.3}{\milli \meter}$) adjacent to the solid surface. In the outer block, the mesh was gradually coarsened using geometric grading. The first cell in this region was matched to the near-wall spacing of approximately $\SI{0.04}{\milli\metre}$, thereby avoiding an abrupt change in cell size at the interface between the two blocks. The cell size then increased progressively with height, giving an average spacing of approximately $\SI{0.25}{\milli\metre}$ over the interval $y=\SI{1}{\milli\metre}$ to $\SI{10}{\milli\metre}$ and a maximum cell size of approximately $\SI{0.65}{\milli\metre}$ near the upper boundary. The hydraulic jump forms at a film height of approximately $\SIrange{3}{4}{\milli\metre}$. Within this region, the local vertical grid spacing varies from approximately $\SI{0.27}{\milli\metre}$ to $\SI{0.33}{\milli\metre}$. The graded mesh therefore retains relatively fine resolution in the vicinity of the jump while allowing progressively coarser cells farther from the wall, where smaller gradients are expected.

\begin{figure}
    \centering
\includegraphics[width=0.9\linewidth]{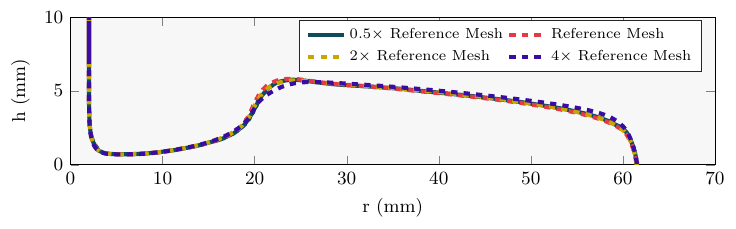}
\caption{Comparison of the simulated film profiles for the WG~10/90 mixture at $Q=\SI{4}{\litre\per\minute}$ using different mesh resolutions. The mesh count was varied by factors of $0.5$, $1$, $2$ and $4$ relative to the reference mesh described above. No significant deviation is observed among the resulting profiles.}
\label{Fig: grid independancy}
\end{figure}

In the $x$-direction, the computational domain was divided into three regions corresponding to the nozzle, the main radial-flow region and the outlet region. A finer uniform grid spacing of $\SI{0.1}{\milli\metre}$ was used within the nozzle region to resolve the incoming jet adequately. A uniform spacing of $\SI{0.25}{\milli\metre}$ was used throughout the main flow and outlet regions. This mesh arrangement provides a computationally efficient simulation by concentrating finer resolution in regions with strong gradients and interfacial dynamics, while employing coarser cells elsewhere without compromising the accuracy of the predicted flow. A grid-independence study was performed using meshes containing $0.5$, $2$ and $4$ times the number of cells in the reference mesh. Here, the reference mesh denotes the mesh configuration described above and adopted for the main simulations. Figure~\ref{Fig: grid independancy} compares the film profiles obtained using the four mesh resolutions. No significant deviation is observed among the profiles, with the predicted upstream film thickness, jump radius and downstream film profile remaining effectively unchanged. Further refinement beyond the reference mesh therefore produces no appreciable change in the solution, confirming that the selected mesh provides a grid-independent result while maintaining a reasonable computational cost. 

\subsubsection{Control parameters}

The simulations were continued until no further variation in the hydraulic-jump position was observed, and the flow was then considered temporally steady. The adaptive time stepping was employed, with an initial and maximum allowable time step of $\Delta t=\SI{1e-3}{\second}$. As a result, the time step was adjusted dynamically to maintain both the global Courant number and the phase-fraction Courant number below $0.5$, i.e.\ $Co_{\max}=Co_{\alpha,\max}=0.5$. These constraints were imposed to ensure numerical stability and accurate advection of the air-liquid interface. The solution fields were written at intervals of $\SI{0.05}{\second}$. The pressure-velocity coupling was performed using the PIMPLE algorithm. Three outer corrector iterations were applied within each time step, with four pressure-correction iterations per outer loop. The momentum predictor was enabled, while no additional non-orthogonal pressure corrections were required because the computational mesh was predominantly orthogonal. The volume-fraction equation was solved using two correction iterations and one subcycle per time step. The multidimensional universal limiter with explicit solution (MULES) correction was enabled, with five limiter iterations, to maintain boundedness of the phase-fraction field. The absolute convergence tolerances for the phase fraction, modified pressure and velocity equations were set to $10^{-8}$, $10^{-7}$ and $10^{-6}$, respectively. A zero relative tolerance was used for all equations, such that each linear system was solved to its prescribed absolute tolerance during every corrector iteration.

\bibliographystyle{jfm}
\bibliography{jfm}

\end{document}